\documentclass[useAMS,usenatbib]{mn2e}
\usepackage{amssymb,epsfig,algorithm,algorithmic}

\title[Accurate SFH estimation in galaxies]{Accurate parameter
  estimation for star formation history in galaxies using SDSS spectra}
\author[J. W. Richards et al.]{Joseph W. Richards$^{1}$\thanks{E-mail:
    jwrichar@stat.cmu.edu}, Peter E. Freeman$^{1}$, Ann B. Lee$^{1}$,
  Chad M. Schafer$^{1}$\\
$^{1}$Department of Statistics, Carnegie Mellon University, 5000
Forbes Avenue, Pittsburgh, PA 15213}

\begin{document}

\date{28 May 2009}

\pagerange{\pageref{firstpage}--\pageref{lastpage}} \pubyear{2009}

\maketitle

\label{firstpage}

\begin{abstract} 
To further our knowledge of the complex physical
process of galaxy formation, it is essential that we characterize the
formation and evolution of
 large databases of galaxies. The spectral synthesis STARLIGHT code of Cid
Fernandes et al. (2004) was designed for this purpose.  Results of
STARLIGHT are highly dependent on the choice of input basis of simple
stellar population (SSP) spectra.  Speed of the code, which uses 
random walks through the parameter space, scales as the square of the number
of basis spectra, making it computationally necessary to choose a
small number of SSPs that are coarsely sampled in age and metallicity. In this
paper, we develop methods based on diffusion map (Lafon \& Lee, 2006)
that, for the first time, choose appropriate bases of prototype SSP
spectra from a large set of SSP spectra designed to approximate the
continuous grid of age and metallicity of SSPs of which galaxies are
truly composed.  We show that our techniques achieve better accuracy
of physical parameter estimation for simulated galaxies.
Specifically, we show that our methods significantly decrease the
age-metallicity degeneracy that is common in galaxy population
synthesis methods.  We analyze a sample of 3046 galaxies in SDSS
DR6 and compare the parameter estimates obtained from different
basis choices.
\end{abstract}


\section{Introduction} 

Determining the physical parameters of large samples of galaxies is
crucial to constrain our knowledge of the complicated physical processes
that govern the formation and evolution of these systems.  Ultimately,
we seek methods that quickly and effectively map observed data from
$\sim 10^6$ galaxies
to sets of parameters that describe the star formation and chemical
histories of these galaxies.  Due to the complexity of the
mechanisms of galaxy evolution, it is critical that we adopt a model
that uses few assumptions about how galaxies evolve.  Moreover, full
utilization of the available theoretical stellar models, avoiding ad
hoc simplifications, is critical to obtaining accurate parameter estimates.

There has been a large amount of work dedicated to estimating the
physical parameters of databases of galaxies using different types of
data.  For example, spectral indices
(\citealt{Worth94,K03,Tre2004,Gal2005,Chen2009}), emission
features (\citealt{Mas91,Leith95}), and full high-resolution,
broad-band spectra
(\citealt{Rei2001,Gla2003,Pan2003,CF2004,CF2005,Mat2006,Ocv2006,Asa2007})
have all been utilized to
estimate the star formation histories (SFHs) and metallicities of
galaxies.  Recently, due to the abundance of high-quality, homogeneous
databases of galaxy spectra such as the Sloan Digital Sky Survey
(SDSS, \citealt{York2000,Str2002}), which provides high-resolution
spectra for hundreds of thousands of
galaxies, the approach of full high-resolution spectral fitting is now
possible to
infer the properties of large populations of galaxies.  Recent work
has shown promise in the application of these methods to SDSS galaxy
spectra (see \citealt{Pan2007} and \citealt{Asa2007} for reviews of
results achieved by two such fitting methods).   However, large-scale
analyses, while possible, are not necessarily accurate, and can be
computationally challenging.  There is much
room for improvement in accuracy of SFH parameter estimation.  It is critical
that we understand the shortcomings of the current models and improve
their accuracy.  At the same time,  to achieve our goal of obtaining 
accurate SFH estimates for a large sample of galaxies, we cannot 
sacrifice the computational efficiency of current methods.

A common technique in the literature, called empirical population
synthesis, is to model galaxies as
mixtures of simple stellar populations (SSPs) with known physical
parameters.  Recent studies that have used this method are,
e.g. \citet{Bica1988}, \citet{Pel1977}, \citet{CF2001}, and
\citet{Mou2004}.   Under this model, galaxy data are treated as linear
combinations of the observed properties of SSPs.  Historically, SSP
parameters and observables were derived from observations of
well-understood stellar systems.  More recent studies have instead
used model-produced SSPs, such as those from the evolutionary
population synthesis models of \citet{BC03}.  The STARLIGHT spectral
fitting code, introduced by \citet{CF2004}, fits observed spectra
with linear combinations of SSPs from the models of \citet{BC03}.  In
\citet{CF2005}, it was shown that SFH parameters of simulated galaxy
spectra could
be recovered by STARLIGHT in the absence of noise.  However, their
simulated spectra
were generated and fit with the \emph{same basis} of 45 SSPs,
rendering the results difficult to generalize to the expected
performance on real galaxies, which can be composed of SSPs on
infinitely fine grids of age and metallicity.

Results of the STARLIGHT fitting code are highly dependent on the
choice of basis of SSP spectra.  The database of SSPs available to us could
theoretically be infinitely large, encompassing all combinations of
age, metallicity, initial mass function, evolutionary prescription,
etc.  Including too many SSPs in our basis causes the code to be
prohibitively slow and the solution to be degenerate due to the inclusion of
SSPs with almost identical spectra.  If, however, we simply
hand-select a few SSPs on a coarse grid of age and metallicity with
which to model our data, then we effectively ignore large
subsets of our SSP parameter space, leading to suboptimal estimates for
fits to databases of galaxies.  Using a hand-selected basis of SSPs
may also lead to the inclusion of multiple SSPs whose spectra are
essentially identical, which can lead to degeneracies.

There is much room for improvement in the
accuracy of SFH estimates by exploring numerical methods of selecting
small bases of SSPs.  In this paper, we propose a
method of choosing a small set of \emph{prototype} spectra from a large
database of SSP spectra based on their observable
quantities.  Our method attempts to capture much of the variability of
a large set of SSPs in a few numerically chosen prototype SSPs.  The
method is based on the diffusion map of
\citet{Coif2006} and \citet{LafonLee2006}, a non-linear technique that
seeks a simple, natural representation of data that are complex and
high-dimensional, such as high-resolution astronomical spectra.

We utilize the \emph{diffusion
  $K$-means} method to numerically find a basis of prototype SSPs with
which we can fit galaxies to accurately estimate their SFHs.  In a
future study, we will consider the problem of computational speed in
methods such as STARLIGHT, which takes $\gtrsim 5$ minutes to fit each
SDSS galaxy spectrum for a basis of size 150, and show how to quickly extend
the SFH estimates obtained by using the methods in this paper
to a large database of galaxies by taking advantage of the
geometry of the manifold on which the data lie.

The applicability of
the methods introduced here extend beyond the problem of SFH
estimation using high-resolution spectra.  Any astronomical task where
observed data (i.e. spectra, photometric data, spectral indices, etc.)
are fit as linear
combinations of observable data from simpler
systems can benefit from intelligent numerical selection of prototypes.

The outline of the paper is as follows.  In \S 2 we describe the
STARLIGHT spectral synthesis code, discuss the drawbacks of SSP bases
employed in the 
literature, and introduce the diffusion $K$-means method of prototype
selection.  We directly compare the basis of prototypes selected by our
method to those used in the literature and those found using other
methods.  In \S 3 we fit several sets of realistic simulated galaxy
spectra using the STARLIGHT code with different bases to directly
compare the ability of each basis to accurately estimate galaxy
parameters.  In \S 4
we fit a set of  SDSS galaxy spectra and analyse physical parameter
estimates obtained by different bases.  We conclude with some
remarks and discussion of future directions in \S 5.

\section{Methods}

\subsection{Spectral synthesis}
\label{specsyn}

\subsubsection{Model and STARLIGHT fitting algorithm}

We adopt the model of galaxy spectra as a mixture of simple systems
introduced in
\citet{CF2004}:
\begin{equation}
  \label{cfmodel}
  M_{\lambda} =
  M_{\lambda_0}\left(\sum_{j=1}^{K}x_j\xi_{j,\lambda}r_{\lambda}\right) \otimes G(v_*,\sigma_*),
\end{equation}
where $M_{\lambda}$ is the model flux in wavelength bin $\lambda$.
The component 
$\xi_{j}$ is the $j$th basis spectrum normalized at wavelength
$\lambda_0$.  The basis of SSP spectra
$\mathbf{\xi}=\{\mathbf{\xi}_1,...,\mathbf{\xi}_K\}$ is chosen before
performing the analysis.  The scalar $x_j \in [0,1]$ is the proportion
of flux
contribution of the $j$th component at $\lambda_0$,
where $\sum_{j=1}^{K}x_j = 1$; the vector $\mathbf{x}$ with components
$x_j, (j=1,...,K)$ is the \emph{population vector} of a galaxy.  These flux
fractions can be converted to mass fractions, $\mu_i$, using the
light-to-mass ratios of each basis spectrum $\xi_j$ at $\lambda_0$.
To describe the SFH of a galaxy, we will use time-binned versions of
$\mathbf{x}$ and $\mathbf{\mu}$, denoted $\mathbf{x}_c$ and
$\mathbf{\mu}_c$, respectively.  Using binned population vectors
lets us avoid degeneracies that can occur in the estimation of the
original population vectors.

Other components of the model in (\ref{cfmodel}) are the scaling parameter
$M_{\lambda_0}$, the reddening term,
$r_{\lambda}=10^{-0.4(A_{\lambda}-A_{\lambda_0})}$, and the Gaussian
convolution, $G(v_*,\sigma_*)$.  The reddening term describes distortion
in the observed spectrum due to foreground dust, and is modeled by
the extinction law of \citet{CCM1989} with $R_V=3.1$.  The Gaussian
convolution, $G(v_*,\sigma_*)$, accounts for movement of stars within
the observed galaxy 
with respect to our line-of-sight, and is parametrized by a central
velocity $v_*$ and dispersion $\sigma_*$.

To fit individual galaxies, we use the STARLIGHT fitting routine
introduced by \citet{CF2005}.  The code uses a Metropolis algorithm
plus simulated annealing to find the minimum of
\begin{equation}
  \label{chisq}
  \chi^2(\mathbf{x},M_{\lambda_0},A_V,v_*,\sigma_*) = \sum_{\lambda=1}^{N_{\lambda}}[(O_{\lambda}-M_{\lambda})w_{\lambda}]^2,
\end{equation}
where $O_{\lambda}$ is the observed flux in the wavelength bin
$\lambda$, $M_{\lambda}$ is the model flux in (\ref{cfmodel}), and
$w_{\lambda}$ is the inverse of the noise in the measurement
$O_{\lambda}$.  The summation is over the $N_{\lambda}$ wavelength
bins in the observed spectrum.  The minimization of (\ref{chisq}) is
performed over $K+4$ parameters: $x_1,...,x_K, M_{\lambda_0}, A_V, v_*$,
and $\sigma_*$, where $K$ is the number of basis spectra in $\mathbf{\xi}$.
The speed of the algorithm scales as $K^2$.  For the
analysis of, e.g. $\simeq 10^6$ galaxy spectra in the SDSS database, it is
crucial to pick a basis with a small number of spectra.

In, for example, \citet{CF2004} and \citet{CF2005}, the STARLIGHT model
(\ref{cfmodel}) and
fitting algorithm are tested using simulations.  The simulated spectra are
generated from the model in (\ref{cfmodel}) using $K=30$ SSPs from
\citet{Tre2003} and $K=45$ SSPs from \citet{BC03}, respectively.  
These authors fit the simulations using the same basis of SSPs that was
used to generate the simulations.  From their
analyses, they conclude that in the absence of noise, the algorithm
accurately recovers the input parameters.  In the presence of noise,
the population vector $\mathbf{x}$ is not recovered, and it is advised
that a time-binned description of the SFH be adopted.  Although their use
of the same basis for both generating and fitting simulated spectra
is an appropriate test that the algorithm works, it
is not a fair assessment of the expected performance of the methods for a
database of real galaxy spectra.  In \S \ref{models} we discuss our
concerns with their simulation assessments.  In \S \ref{basis} we
present a new suite of methods that we test with several sets of
simulated galaxy spectra.  These simulations, which are described in
\S \ref{simulations}, are designed to
be more representative of a large database of galaxy spectra, such as SDSS.

\subsubsection{Stellar population models}
\label{models}

In reality, the population of all observed galaxies is not constrained
to be mixtures of a small number of SSPs on a discrete grid of age and
metallicity.  A more
physically accurate representation of galaxies is as mixtures of an
infinitely large basis of SSPs on a 
continuous grid of age and metallicity and, depending on the
complexity of the underlying physics, a possibly infinite
grid of prescriptions for initial mass function and evolutionary
track.  In our simulations in \S \ref{simulations} we simulate galaxies
from a large database of $\sim 10^3$ SSPs on a fine grid of age and
metallicity to
determine whether we can accurately describe their
SFHs, metallicities, and kinematic parameters using an
appropriately chosen, computationally tractable basis of $K=45$ or 150
spectra.

We adopt a SSP database containing 1278 spectra.  This set of spectra
is used to both choose an
appropriate basis (\S \ref{basis}) and generate simulated spectra (\S
\ref{simulations}).  The database is meant to represent the large
population of SSPs from which observed galaxies can be composed.  We
use the spectra from the models of \citet{BC03},
computed for a \citet{Cha2003} IMF, `Padova 1994' evolutionary tracks
(\citealt{Alo1993,Bre1993,Fag1994a,Fag1994b,Gir1996}), and STELIB library
(\citealt{LeB2003}).  Theoretically, we could also vary the IMF and
evolutionary tracks across spectra and use the same methods developed in
\S \ref{basis}, but for simplicity decide to fix them in this study.
The SSPs in our database are generated on a grid of
213 approximately evenly log-spaced time bins
from 0 to 18 Gyrs and 6 different metallicities: $Z$ = 0.0001, 0.0004,
0.004, 0.008, 0.02 and 0.05, where $Z$ is the fraction of the mass of a
star composed of metals ($Z_{\odot}=0.02$).  This grid is designed to
approximate a continuous variation across age and $Z$.  The coarseness
in the current $Z$ grid is due to limitations in the current stellar
population models.  Again, a finer
grid of age and $Z$ can be implemented with the techniques that we
develop in \S \ref{basis}.

\subsection{Choosing an appropriate basis of prototype spectra} 
\label{basis}

In previous studies using the STARLIGHT code, researchers
hand-selected sets of SSP spectra to use in the model fitting.
According to \citet{CF2004}, ``the elements of the
base should span the range of spectral properties observed in the
sample galaxies and provide enough resolution in age and metallicity
to address the desired scientific questions.''  In this same paper,
they claim that their basis of $K=30$ SSPs adequately recovers
parameters of age and metallicity in their simulations.  However, as
discussed in \S \ref{specsyn}, these simulations are unrealistic since
they employ the same basis to both generate and fit the simulated spectra.

In this section we present methods to numerically determine a basis
$\mathbf{\xi}$ of a small number of \emph{prototype spectra}, $\xi_j$, from
a large set of SSP spectra.  The resultant basis of prototype spectra
will be designed to capture a large proportion of the variation of the set of
SSPs, which is chosen to span the range of observed spectral
properties of a data set of galaxies at a high resolution in age
and metallicity, such as the database of 1278 SSPs described in \S
\ref{models}.  The numerically chosen basis of
prototype spectra, $\mathbf{\xi}$, should recover physical
parameters better than any hand-chosen basis of the same size because
hand-selected bases ignore subsets of the parameter space and may
cause the solution to be degenerate by including multiple SSPs with nearly
identical spectra. In \S\ref{simulations} we show that
the bases computed from our numerical methods outperform hand-selected
bases used in the literature in parameter estimation for
simulated galaxies.

\subsubsection{Diffusion $K$-means}

Diffusion map is a popular new technique for data
parametrization and dimensionality reduction that has been developed
by \citet{Coif2006} and \citet{LafonLee2006}, and was recently used by
\citet{Rich2009} to analyse SDSS spectra and \citet{Free2009} to
estimate photometric redshifts.  As shown in these latter works,
diffusion map has the powerful ability to uncover simple structure in
complicated, high-dimensional data.  These analyses demonstrate
that this simple structure can be used to make precise statistical
inferences about the data.

In the problem at hand, we begin with a set $\mathbf{B}$ of
high-resolution broad-band SSP
spectra $\mathbf{b}_{i}, i=1,...,N$, where $N$ can be very
large and each  $\mathbf{b}_{i}$ has flux measurements in $p \gtrsim
10^3$ wavelength bins.  From this set, our goal is to derive a small basis of $K$ prototype
spectra, $\mathbf{\xi} =
\{\mathbf{\xi}_{1},...,\mathbf{\xi}_{K}\}$, each of length $p$, that
captures most of the variability in $\mathbf{B}$.  In what follows, we
briefly review the basics of the diffusion map technique and then show
how it can be used to find an appropriate basis $\mathbf{\xi}$.

The main idea of the diffusion map is that it finds a parametrization
of a data set that preserves the connectivity
of data points in a way that is dependent only on
the relationship of each datum to points in a local neighborhood.  By preserving
the local interactions in a data set, diffusion map is able to learn,
 e.g. a natural parametrization of the of the spiral in Fig. 1 of
\citet{Rich2009} where other methods, such as principal components
analysis, fail.

Starting with our set of SSP spectra, $\mathbf{B}$, in units of
$L_{\odot}$/\AA  ~per initial $M_{\odot}$, we normalize the
spectra at $\lambda_0$ to ensure that we  are not confused by absolute scale
differences between individual spectra.  We retain the scaling factors
$\gamma_{i,\lambda_0}$ for later use to convert back to the original
units.  Next, we select a discrepancy measure between SSPs, i.e.
\begin{equation}
  \label{distance}
  s(\mathbf{b}_i,\mathbf{b}_j) = \sqrt{\sum_{\lambda=1}^{p}(b_{i,\lambda}-b_{j,\lambda})^2},
\end{equation}
which is the Euclidean distance between two SSP spectra.  
The specific choice of $s(\mathbf{b}_i,\mathbf{b}_j)$ is often not
crucial to the 
diffusion map; any reasonable measure of discrepancy will return
similar results.  It is important to note here the flexibility of the
diffusion map method.  For each problem, we can define a different
discrepancy measure based on the type of data available to us and the
types of information we want to incorporate. For example, we could
define our measure $s$ by weighting more heavily those regions around
absorption lines which are highly correlated with stellar age and
metallicity.  Contrast this flexibility with the rigidity of principal
components analysis, which 
is only based on the correlation structure of the data.

From our discrepancy measure in (\ref{distance}), we construct a
weighted graph on our data set $\mathbf{B}$, where the nodes are the
individual SSP spectra and the weights on the edges of the graph
reflect the amount of similarity between pairwise spectra.  We define
the weights as
\begin{equation}
  \label{weights}
  w(\mathbf{b}_i,\mathbf{b}_j) = \exp\left(-\frac{s(\mathbf{b}_i,\mathbf{b}_j)^2}{\epsilon}\right)
\end{equation}
where $\epsilon$ should be small enough such that
$w(\mathbf{b}_i,\mathbf{b}_j)\approx 0$ unless $\mathbf{b}_i$ and
$\mathbf{b}_j$ are similar, but large enough so that the entire graph
is connected (i.e. every SSP has at least one non-zero weight to
another SSP).  In practice, we choose $\epsilon$ that
produces a basis that achieves the best fits to galaxy spectra (see
\S\ref{simulations}).
By normalizing the rows of the weight matrix, $\mathbf{W}$, to sum to
unity, we can define a Markov random walk on the graph where the
probability of going from $\mathbf{b}_i$ to $\mathbf{b}_j$ in one step
is $p_1(\mathbf{b}_i,\mathbf{b}_j) =
w(\mathbf{b}_i,\mathbf{b}_j)/\sum_kw(\mathbf{b}_i,\mathbf{b}_k)$.  We
store all of these one-step probabilities in a matrix $\mathbf{P}$.

By the theory of Markov chains, the $t$-step transition probabilities
are defined by $\mathbf{P}^t$.  In
the previous analyses in \citet{Rich2009} and \citet{Free2009}, the
authors fixed $t$.  In these previous papers, the choice of $t$ was
unimportant because the task was linear regression on the diffusion
coordinates, in which the multipliers
$\lambda_j^t$ of the diffusion map were absorbed into the linear
regression coefficients $\beta_j$.  
In the present work, our task is to determine a set of $K$
prototype spectra.  The tool we use is $K$-means clustering in
diffusion space, where
results are dependent upon intra-point Euclidean distances of
the diffusion map, which in turn are dependent on $t$.  

In this paper we introduce
a diffusion operator that considers \emph{all} scales $t=1,2,3,...$
simultaneously.  We define the \emph{multi-scale} diffusion map as 
\begin{equation}
  \label{dmap2}
  \mathbf{\Psi} : \mathbf{b}_i \mapsto
  \left[\frac{\lambda_1}{1-\lambda_1} \psi_1(\mathbf{b}_i),\frac{\lambda_2}{1-\lambda_2} \psi_2(\mathbf{b}_i),\cdots,\frac{\lambda_m}{1-\lambda_m}\psi_m(\mathbf{b}_i)\right].
\end{equation}
where $\psi_j$ and $\lambda_j$ are the right eigenvectors and eigenvalues of
$\mathbf{P}$, respectively, in a bi-orthogonal decomposition.
Euclidean distance in the $m$-dimensional
space described by equation (\ref{dmap2}) approximates the multi-scale
diffusion distance, a distance measure that utilizes the
geometry of the data set by simultaneously considering all possible
paths between any two data points at all scales $t$, in the Markov
random walk constructed above.
An advantage to using the multi-scale diffusion map, besides
elimination of the tuning parameter $t$, is that it gives a more
robust description of the structure of the data by considering the
propagation of local interactions of data points through all
scales.

We also have three other tuning
parameters: $\epsilon$, $m$ (dimensionality of the diffusion map), and
$K$ (number of prototypes) over which to optimize the fits to galaxy spectra.
For simplicity, we fix $m$ by choosing
the dimensionality of the diffusion map to coincide with a 95\%
drop-off in the eigenvalue multipliers in the multi-scale diffusion
map.  This is reasonable because most of the information about the
structure of the data will be in the few dimensions with large
eigenvalue multipliers.  Simulations show that this choice of $m$ produces
results comparable to optimizing over a large grid of $m$.  For the
parameter $K$, fits should improve as the basis gets larger. However,
we want to
keep $K$ small both for computational reasons and to avoid the
occurrence of nearly identical spectra in our basis, which can cause
degeneracies in our fits.  In practice, we 
choose $K$=45 or 150 to directly compare our results with the bases used
by \citet{CF2005} (CF05) and \citet{Asa2007} (Asa07).  This leaves us with only one
tuning parameter, $\epsilon$, over which to optimize the diffusion
$K$-means basis in fits to galaxy spectra.

Finally, we determine a set of $K$ prototype spectra by performing
$K$-means clustering in the $m$-dimensional diffusion map
representation (\ref{dmap2}) of the SSP spectra, $\mathbf{B}$.  The $K$-means
algorithm is a standard machine learning method used to cluster $N$
data points into $K$
groups.  It works by minimizing the total intra-cluster variance,
\begin{equation}
  \label{kmeans}
  V = \sum_{j=1}^K \sum_{i: \mathbf{b}_i\in S_j} ||\mathbf{\Psi}(\mathbf{b}_i)-\mathbf{c}(S_j)||_2^2
\end{equation}
where $S_j$ is the set of points in the $j$th cluster and $\mathbf{c}(S_j)$ is
the $m$-dimensional geometric centroid of the set $S_j$.  As described
in \S 3.2 of \citet{LafonLee2006}, for diffusion maps 
\begin{equation}
  \label{centroid}
  \mathbf{c}(S_j) = \sum_{i:\mathbf{b}_i \in S_j} \frac{\phi_0(\mathbf{b}_i)\mathbf{\Psi}(\mathbf{b}_i)}{\sum_{\mathbf{b}_i \in
      S_j}\phi_0(\mathbf{b}_i)} ,
\end{equation}
where $\phi_0$ is the trivial left eigenvector of $\mathbf{P}$.  The
$K$-means algorithm begins with an initial partition of the data and
alternately computes cluster centroids and reallocates points to the
cluster with the nearest centroid.  The algorithm stops when no points
are allocated differently in two consecutive iterations.  The final
centroids define the $K$ prototypes to be used in our basis.

\begin{figure}
  \centering
  \epsfig{figure=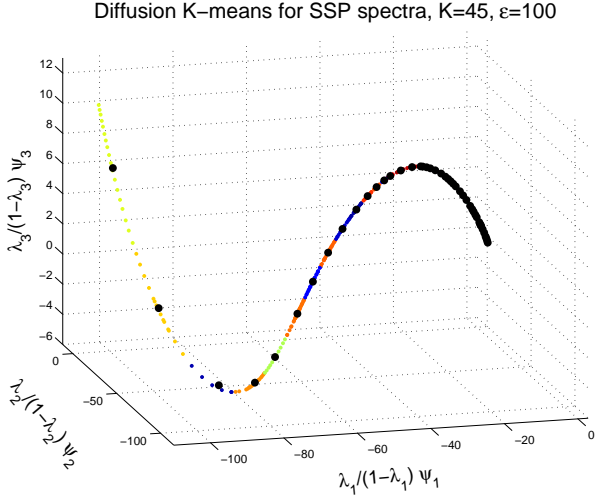,width=80mm}
  \caption{\emph{Diffusion $K$-means for SSP spectra.}  Representation of
    1278 SSP spectra in 3-dimensional diffusion space.  Large black
    dots denote the $K=45$ centroids.  Individual SSPs are coloured by
    cluster membership.  The SSPs reside on a simple, low-dimensional
    manifold which is captured by the $K$ prototypes.}
  \label{sspdmap}
\end{figure}

In Fig. \ref{sspdmap}, we plot the mapping of 1278 SSP spectra
into the $m=3$ dimensional diffusion space described by
(\ref{dmap2}).  The large black dots denote the $K$-means centroids
for $K=45$.  Individual SSPs are coloured by cluster membership.  The
$K$ prototypes capture the variability of the SSP spectra along a
low-dimensional manifold in diffusion space.  Note that the density of
prototypes varies along different parts of the manifold.  This is due
both to the local complexity of the manifold and the sampling of the
original base of SSPs.  This latter effect, that we tend to obtain
more prototypes in areas where there are more SSPs in our basis, can
easily be corrected for by using a weighted $K$-means method,
described in \S 2.2.3, that
accounts for varying density of SSPs along the manifold.

The last step of the diffusion $K$-means algorithm is to determine the
$K$ protospectra and their properties.  First, the normalization constant
for the $j$th prototype spectrum is
\begin{equation}
  \widehat{\gamma}_j = \sum_{\mathbf{b}_i \in S_j} \frac{\phi_0(\mathbf{b}_i)\gamma_i(\mathbf{b}_i)}{\sum_{\mathbf{b}_i \in S_j}\phi_0(\mathbf{b}_i)}
\end{equation}
and the $j$th protospectrum is
\begin{equation}
  \label{eqn:protospec}
  \mathbf{\xi}_j = \widehat{\gamma}_j\sum_{\mathbf{b}_i \in S_j} \frac{\phi_0(\mathbf{b}_i)\mathbf{b}_i}{\sum_{\mathbf{b}_i \in S_j}\phi_0(\mathbf{b}_i)}
\end{equation}
Similarly, the age, metallicity, and ``stellar-mass fraction''---the
percent of the initial stellar mass still in stars---of
prototype $j$ are naturally defined as
\begin{eqnarray}
  \label{eqn:protot}
  t(\mathbf{\xi}_j) &=& \sum_{\mathbf{b}_i \in S_j} \frac{\phi_0(\mathbf{b}_i)t(\mathbf{b}_i)}{\sum_{\mathbf{b}_i \in S_j}\phi_0(\mathbf{b}_i)}\\
  \label{eqn:protoZ}
  Z(\mathbf{\xi}_j) &=& \sum_{\mathbf{b}_i \in S_j} \frac{\phi_0(\mathbf{b}_i)Z(\mathbf{b}_i)}{\sum_{\mathbf{b}_i \in S_j}\phi_0(\mathbf{b}_i)}\\
  \label{eqn:protosmf}
  f_{\star}(\mathbf{\xi}_j) &=& \sum_{\mathbf{b}_i \in S_j} \frac{\phi_0(\mathbf{b}_i)f_{\star}(\mathbf{b}_i)}{\sum_{\mathbf{b}_i \in S_j}\phi_0(\mathbf{b}_i)}
\end{eqnarray}
where $t(\mathbf{b}_i),Z(\mathbf{b}_i)$, and $f_{\star}(\mathbf{b}_i)$ are
the age, metallicity and stellar-mass fraction of the $i$th SSP,
respectively.  Equations (\ref{eqn:protot})-(\ref{eqn:protosmf}) are
natural definitions of the prototype parameters in the sense that the
fit of this basis to a galaxy would yield exactly the same parameter estimates
as the equivalent mixture of the original set of SSPs, $\mathbf{B}$.

\begin{figure}
  \centering
  \epsfig{figure=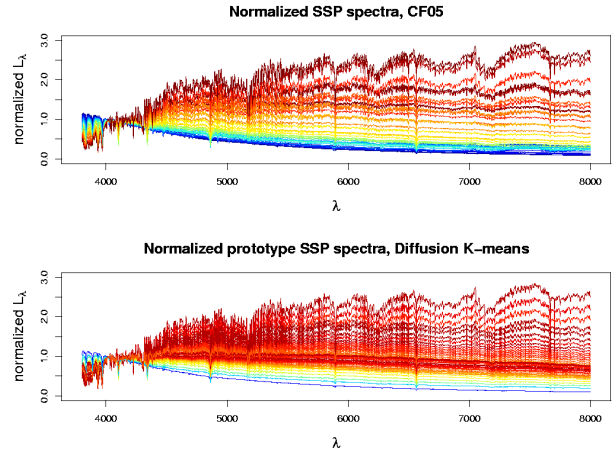,width=80mm}
  \caption{\emph{Basis spectra for CF05 and Diffusion $K$-means,
      coloured by $\log t$.}  All
  spectra are normalized to 1 at $\lambda_0=4020$ \AA.  The diffusion
  $K$-means basis covers the spectral range in gradual increments
  while the CF05 basis includes many young spectra with similar
  spectral properties and sparsely covers the spectral range of older
  populations.}
  \label{protot}
\end{figure}

In Fig. \ref{protot} we plot the 45 SSP spectra in CF05 and the 45
prototype spectra found by diffusion
$K$-means.  All spectra are normalized to 1 at $\lambda_0$=4020\AA~
and are coloured by log age.  The diffusion $K$-means prototypes
spread themselves evenly over the range of spectral
profiles, capturing a gradual trend from young to old spectra. On the
other hand, the
CF05 basis includes many similar spectra from younger
populations and sparsely covers the range of spectral profiles of
older populations.

See Algorithm 1 for a summary of the diffusion $K$-means algorithm.
Sample Matlab and R code is available on the web at {\tt http://www.stat.cmu.edu/\~{}annlee/software.htm}.  The diffusionMap R package is available at {\tt http://cran.r-project.org/}.

\begin{algorithm}
\caption{Diffusion $K$-means}
\label{alg:diffkmeans}
\begin{algorithmic}[1]
\STATE Normalize SSP spectra, $\mathbf{B}$, at $\lambda_0$, fix $\epsilon>0$
\STATE Compute similarity matrix, $\mathbf{S}$ (\ref{distance}) and
weight matrix $\mathbf{W}$ (\ref{weights})
\STATE Compute $\mathbf{P}$: $p_1(\mathbf{b}_i,\mathbf{b}_j) =
w(\mathbf{b}_i,\mathbf{b}_j)/\sum_kw(\mathbf{b}_i,\mathbf{b}_k)$
\STATE Decompose $\mathbf{P}$: $p_1(\mathbf{b}_i,\mathbf{b}_j) =
\sum_{k}\lambda_k \psi_k(\mathbf{b}_i)\phi_k(\mathbf{b}_j)$
\STATE Project $\mathbf{B}$ to the $m$-dimensional diffusion map, $\mathbf{\Psi}$
(\ref{dmap2})
\STATE {\bf $K$-means:}
\STATE \hspace{.2in} Set $k$=1.  Fix $K$.
\STATE \hspace{.2in} Randomly partition data into sets $S_1^{(0)},...,S_K^{(0)}$
\STATE \hspace{.2in} {\bf while} $S_j^{(k-1)} \ne S_j^{(k)}$ for some $j$
\STATE \hspace{.5in} $k = k+1$
\STATE \hspace{.5in} Compute cluster centroids, $\mathbf{c}(S_j^{(k)})$
(\ref{centroid})
\STATE \hspace{.5in} Partition the data so that \begin{displaymath}\mathbf{b}_i \in
\textrm{argmin}_{S_j^{(k)}} ||\mathbf{\Psi}(\mathbf{b}_i)-\mathbf{c}(S_j^{(k)}))||_2^2\end{displaymath}
\STATE \hspace{.2in} {\bf end while}
\STATE {\bf return} $K$ protospectra (\ref{eqn:protospec}) and
corresponding parameters (\ref{eqn:protot}-\ref{eqn:protosmf})
\end{algorithmic}
\end{algorithm}

\subsubsection{Other $K$-means algorithms}

In \S \ref{simulations}, we compare the fits obtained by the diffusion
$K$-means basis to those found by two other numerical methods: 
  principal components (PC) $K$-means and standard $K$-means.  Each of
  these two methods is similar to diffusion $K$-means, but both have
  critical drawbacks.

\emph{PC $K$-means} works similarly to diffusion $K$-means (Algorithm 1)
except that $K$-means is performed on the projection of the normalized
SSP spectra into \emph{principal components} (PC) space, not diffusion
space.  For an example of the application of principal components
analysis to SSP spectra see \citet{Ron1999}.  The main drawback to PC
$K$-means is its assumption that the SSP spectra lie on a linear
subspace of the original $p\gtrsim 10^3$ dimensional space.
If the SSPs actually lie on a non-linear manifold, then
the $K$ prototypes may poorly capture the intrinsic variation of the
original SSPs because the non-linear structure will have been
inappropriately collapsed on to a linear space by the principal
components projection.

\emph{Standard $K$-means} is also similar to Algorithm 1, except that
$K$-means is performed in the original $N_{\lambda}\simeq 10^3$
dimensional space; i.e., no reduction in dimensionality is done before
running $K$-means.  There are two obvious drawbacks to this procedure.
First, the algorithm generally is slow because distance computations are
cumbersome in high dimensions and $K$-means usually takes more
iterations to converge.    For comparison, the dimensionality of the
spaces used by diffusion and PC $K$-means are each $\lesssim 10$, a
factor of 100 smaller than $p$.
Second, and more importantly, appropriate prototype spectra are
difficult to find by standard $K$-means because Euclidean distances, 
used by $K$-means to define clusters, are only physically meaningful
over short distances.  Diffusion $K$-means avoids this problem by
clustering in diffusion space, in which Euclidean distance
approximates diffusion distance, a measure that has physical meaning
on all scales.  The result is that standard $K$-means inappropriately
relates SSPs that are not physically similar.

\begin{figure}
  \centering
  \epsfig{figure=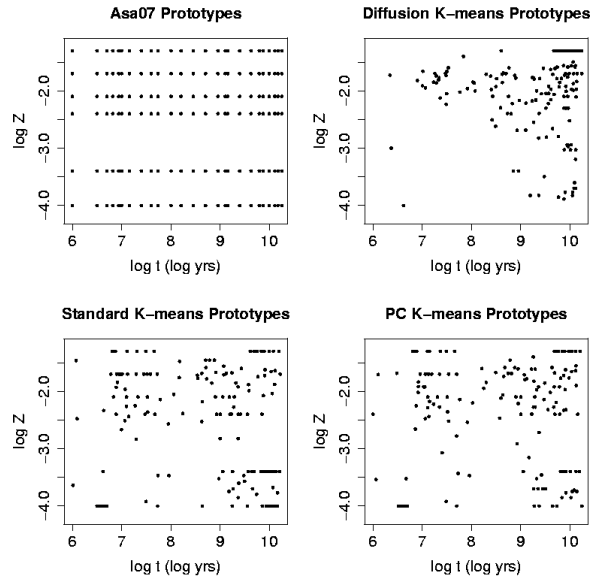,width=80mm}
  \caption{\emph{Plot of $\log Z$ versus $\log t$ for $K$=150
prototypes in Asa07 and $K$-means bases.}
There is a high density of diffusion $K$-means prototypes along a curve that
encompasses large ages and metallicities.  Due to the complicated
manner in which those SSP spectra vary, many more prototypes are
needed to span that spectral range.
Contrast this to the regular grid of Asa07 prototypes.}
  \label{prototZ}
\end{figure}

In Fig. \ref{prototZ}, we plot $\log Z$ versus $\log t$ for $K$=150
prototypes in Asa07 and diffusion, standard, and PC $K$-means.
Notably, diffusion $K$-means finds a much higher density of prototypes
with high $\log Z$ or high $\log t$, reflecting the complicated manner
in which SSPs with those properties vary with respect to $Z$ and $t$.
At the other extreme, the Asa07 prototypes reside on a regular grid,
and thus include many prototype spectra that are essentially identical and also
exclude prototypes that have unique spectral properties.  The standard
and PC $K$-means prototypes also estimate more high $\log Z$ and high
$\log t$ prototypes than a regular grid.

\subsubsection{Incorporating other information}

The methods introduced above use only the observable properties of
SSPs to choose representative prototypes.  It may be the case that we
want to incorporate other \emph{a priori} information that we
have about the SSPs and their relationship with the galaxies we are
fitting.  For instance, we might know that a SSP with a particular
age and metallicity is generally found in the types
of galaxies we are trying to fit, and hence will want to include in
our basis a prototype with characteristics closely matching those of
this SSP.  This information can easily be incorporated with the
framework introduced above by defining an \emph{a priori} weight,
$w_i \ge 0$ for each SSP in our database, where higher weights signify
more importance of the SSP.

By modifying the definition of the diffusion map geometric centroid
(\ref{centroid}) to be
\begin{equation}
  \label{centroid1}
  \mathbf{c}_w(S_j) = \sum_{i:\mathbf{b}_i \in S_j} \frac{w_i\phi_0(\mathbf{b}_i)\mathbf{\Psi}(\mathbf{b}_i)}{\sum_{\mathbf{b}_i \in
      S_j}w_i\phi_0(\mathbf{b}_i)} ,
\end{equation}
and altering the $K$-means algorithm to minimize
\begin{equation}
  \label{kmeans1}
  V_w = \sum_{j=1}^K \sum_{i: \mathbf{b}_i\in S_j} w_i ||\mathbf{\Psi}(\mathbf{b}_i)-\mathbf{c}_w(S_j)||_2^2
\end{equation}
instead of $V$ in (\ref{kmeans}), we choose a basis that reflects
both the numerical observable properties of the SSPs and
our \emph{a priori} knowledge.

This weighted $K$-means algorithm can
be used to correct for the tendency of the prototype SSPs to
depend on the specific set of SSPs, $\mathbf{B}$, chosen.  If, for example, we
define $w_i$ as the inverse of the local density of SSP $i$, then the
prototype SSPs in Fig. \ref{sspdmap} would more 
uniformly cover the manifold of SSPs in diffusion space, and not tend
to more heavily sample denser regions of the SSP manifold.  In this
manner, the prototype spectra will only depend on the spectral
properties, and not the particular sampling of the set of SSPs.

\section{Simulations}
\label{simulations}

We use simulations to assess the performance of the diffusion $K$-means
method for selection of a basis of prototype SSP spectra compared to bases
derived by other $K$-means methods and the bases of $K=45$ used by
CF05 and $K=150$ used by Asa07.  We analyse the performance of these
methods on physically-motivated simulated spectra from two different
prescriptions.  All simulated galaxy spectra are generated
as mixtures of a fine grid of 1278 SSP spectra from the model of
\citet{BC03}, as described in \S \ref{models}.  We construct the
simulated spectra with a fine grid of SSPs because we 
expect real galaxies to consist of sub-populations of stars on a continuous
grid of age and $Z$.  This approach differs significantly from the simulation
prescriptions of, for example, \citet{CF2004} and \citet{CF2005} as
discussed in \S \ref{models}.

 Like \citet{CF2005}, we simulate spectra over the range
 3650-8000~\AA.  We bin all spectra to one measurement every 1~\AA.   For each
simulated galaxy, we independently sample the reddening parameter,
$A_V$, and velocity
dispersion, $\sigma_*$, from a realistic distribution based on fits to a
large sample of SDSS galaxies by the STARLIGHT code that was performed by the
SEAGal (Semi Empirical Analysis of Galaxies) Group and is available on
the web at {\tt www.starlight.ufsc.br}.  We add Gaussian
noise following the
normalized mean SDSS error spectrum with specified S/N at $\lambda_0 =
4020$~\AA.  We mask out any spectral bins around emission lines,
as done in \citet{CF2005}.

To choose the optimal value of $\epsilon$ for the diffusion $K$-means basis,
we fit simulations to each basis from a grid of $\epsilon$ values and
find the basis for which the fits achieve the smallest median $\chi^2$
value, (\ref{chisq}).
Although we choose the optimal $\epsilon$ by minimizing $\chi^2$, our
ultimate goal is accurate estimation of the SFH and physical
parameters for each galaxy.  However, we find that bases that
achieve small $\chi^2$ values tend to have small errors in their physical parameter
estimates.

To compare results across different bases, we compare the errors in
physical parameter and SFH estimation.  Our ultimate goal is to find the
basis that most accurately describes the star formation and metallicity
history of each galaxy.
In this study, our main parameters of interest are
\begin{eqnarray}
\langle \log t \rangle_L &=& \sum_{j=1}^K x_j \log (t(\xi_j)), \\
\log \langle Z \rangle_L &=& \log\left(\sum_{j=1}^K x_j Z(\xi_j)\right),
\end{eqnarray}
$A_V$, and $\sigma_*$.  To test the ability of different bases
to estimate these parameters, we 
compute the mean square error (MSE) of the estimates across the 
simulated galaxies.  To characterize the ability to recover
SFH, we compute condensed population vectors by binning both
$\mathbf{x}$ and $\mathbf{\mu}$ into 5 logarithmically spaced time bins and
then comparing these true condensed vectors,
$\mathbf{x}_c$ and $\mathbf{\mu}_c$ to the estimated condensed
vectors, $\mathbf{\widehat{x}}_c$ and $\mathbf{\widehat{\mu}}_c$.
Other techniques have proposed different schemes for compressing the
SFH information retrieved by spectral fitting
(see \citealt{Asa2007} for an overview of the different compression
methods employed in current spectral synthesis codes).  Our choice of
5 age bins is an attempt to avoid the degeneracies that are likely if
the population vector were gridded into very fine age bins while still
retaining a sufficient number of age bins to detect significant recent
starburst
events.  Admittedly, the choice of 5 age bins is ad hoc.  However,
because we are only interested in comparing the ability of different
bases of prototype spectra to recover SFH, any reasonable
parametrization of the SFH will allow us to perform this analysis.

\subsection{Exponentially decreasing SFH}
\label{sim:exp}

In the first set of simulations, we choose a SFH with exponentially
decaying star formation rate (SFR): SFR $\propto \exp(\gamma t)$,
where $t$ is the age of
the SSP in Gyr and we use SSPs up to a maximum of 14 Gyrs.  Like
\citet{toj2007}, we set $\gamma = 0.3
\textrm{Gyr}^{-1}$ to have a SFH that is neither dominated by recent
star formation nor too similar to a single old burst galaxy.  In
our simulations, we consider {\it specific} SFR (SSFR) as defined in
\citet{Asa2007} by normalizing all simulated spectra to have the same
mass.  This is reasonable because in practice we are interested in how
the SFH varies across many galaxies, and therefore wish to remove the
absolute mass-scale dependence of each object's SFR.  We have SSPs
from 196 time bins where each bin's metallicity is
randomly chosen from 6 different values. We test our methods on 50 simulated
spectra.  Across different spectra the SFH is the same, while
$\log\langle Z_*\rangle_L, A_V, \sigma_*$ and the random noise all
vary.  Each spectrum has S/N=10 at $\lambda_0$. 

\subsubsection{Results}

The following bases of prototype spectra are used to fit each simulated
spectrum: 1) the CF05 basis, 2) diffusion $K$-means, 3) PC $K$-means, and 4)
standard $K$-means.  Each basis contains $K=45$ prototype spectra.
For diffusion $K$-means, we perform fits over a grid of values of
$\epsilon$ to study how parameter estimates vary as a function of $\epsilon$.

\begin{figure}
  \centering
  \epsfig{figure=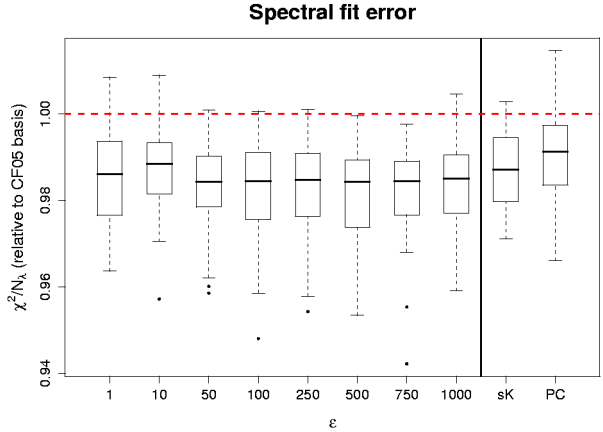,width=80mm}
  \caption{\emph{Spectral fit errors to simulated spectra with
    exponentially decreasing SFH, $\gamma=0.3$.}  Errors are plotted
    relative to the error obtained by the CF05
    basis (dashed line).  For each value of $\epsilon$, diffusion
    $K$-means outperforms the CF05 basis, as do
 both standard $K$-means (sK) and PC $K$-means (PC).}
  \label{toj:err}
\end{figure}

In Fig. \ref{toj:err} we show the error in spectral fits
using diffusion $K$-means for different values of $\epsilon$.  Also
plotted is the error
in the standard $K$-means (sK) and PC $K$-means
(PC) bases.  All errors are plotted as the ratio of the $\chi^2$ statistic
to the $\chi^2$ from the CF05 basis fit.  In the vast majority of
the simulated spectra, all three of the techniques considered produce
better spectral fits than the
CF05 basis.  There are no significant differences between
spectral fit errors for the diffusion $K$-means bases as $\epsilon$
varies.  The minimum median $\chi^2$ for diffusion $K$-means is
achieved at $\epsilon=750$. 

\begin{figure}
  \centering
  \epsfig{figure=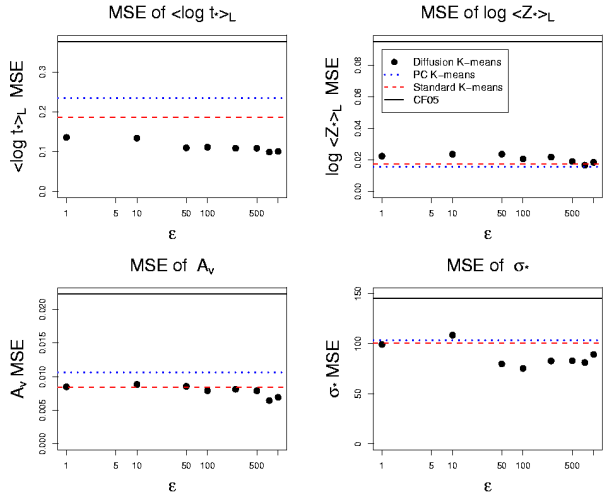,width=80mm}
  \caption{\emph{MSE of four parameter estimates,
    for simulated spectra with exponentially-decaying SFH,
    $\gamma=0.3$.}
    MSE is computed across 50 simulated spectra with S/N=10.
    Parameter estimates from the diffusion $K$-means bases (solid
    dots) are
    plotted along with those from the CF05 basis (solid
    line), standard $K$-means basis (dashed line), and PC $K$-means basis
    (dotted line).  The diffusion $K$-means basis generally
    outperforms the other bases considered; CF05 estimates poorly for
    all 4 parameters.}
  \label{toj:mse}
\end{figure}

We plot the MSE of our estimates for
$\langle\log t_* \rangle_L, \log\langle Z_* \rangle_L, A_V$, and
$\sigma_*$ in Fig. \ref{toj:mse}.  For each parameter, the CF05
basis predictions are the worst in terms of MSE.  The performance of
the diffusion $K$-means basis is not very dependent on the particular
choice of $\epsilon$.  This is a nice property, as any reasonable
choice of $\epsilon$ will give us good results, and we need not spend
computing time searching for an optimal value of $\epsilon$.
Diffusion $K$-means bases tend to find the best estimates of
$\langle\log t_* \rangle_L$, $A_V$, and $\sigma_*$, and performs
competitively with the other $K$-means bases for $\log\langle Z_*
\rangle_L$ estimation.

\begin{figure}
  \centering
  \epsfig{figure=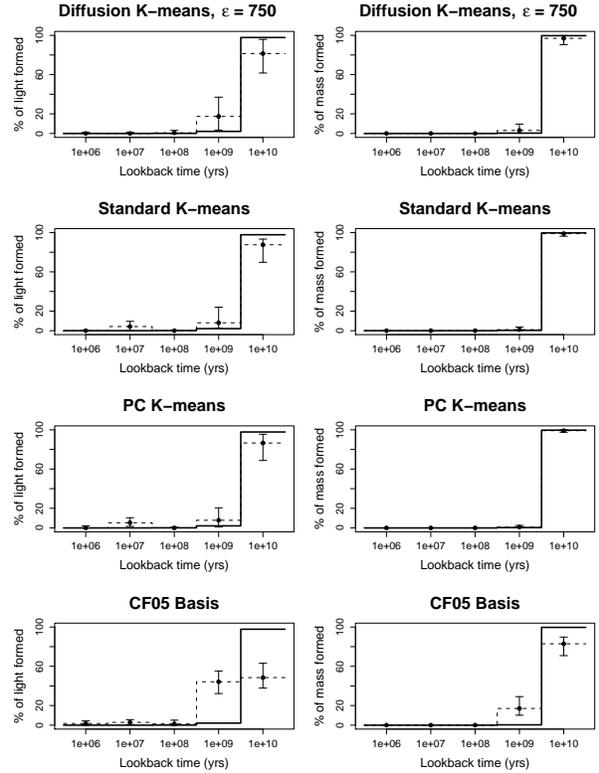,width=80mm}
  \caption{\emph{Estimated star formation histories for simulated spectra with
      exponentially-decaying SFH, $\gamma=0.3$.}
    The estimated SFHs for 4
    different bases are plotted as dots with dashed lines to denote the median
    estimated contribution of 
    galaxy light (left column) and galaxy mass (right column) formed
    in each time bin for fits to 50 simulations. 
    Error bars denote the 90\% prediction intervals for the fits.  The true,
    underlying model, SFR $\propto \exp(\gamma t)$, is plotted as a solid
  line.  The CF05 basis performs much worse than the $K$-means bases,
  confusing old populations for younger ones.}
  \label{toj:sfh1}
\end{figure}

We compare the condensed SFH estimates across bases.
Plots of the recovered
$\mathbf{\widehat{x}}_c$ and $\mathbf{\widehat{\mu}}_c$ vectors in
Fig. \ref{toj:sfh1} show that the 
CF05 basis tends to significantly underestimate the contributions of the
oldest population while overestimating the contribution of the
second oldest population.  All three $K$-means bases find estimates
that are closer to the true $\mathbf{\widehat{x}}_c$, but tend
to underestimate the 
contribution of the oldest $\mathbf{\widehat{x}}_c$ component.
All three $K$-means bases find very accurate estimates of
$\mathbf{\widehat{\mu}}_c$.
The bias in $\mathbf{\widehat{x}}_c$ estimation occurs for at least
two reasons.  First, since the simulations
are generated from SSPs on a fine age grid, it is intrinsically
difficult to estimate all of the contributions of the
subpopulations.  Second, due to the exponential form of the SFR, older
populations are the most important in the SFH, making it easy to underestimate their
importance and almost impossible to overestimate their contribution.

For this simple prescription of galaxy evolution, bases constructed using
$K$-means methods find significantly better estimates of physical
parameters and SFH than the CF05 basis.  Next, we test our methods on
a slightly more complicated set of simulations and also compare bases
of size 45 and 150 to characterize the advantage, if any, of using a
larger basis and whether this advantage is worth the extra expenditure
in computation time.

\subsection{Exponentially decreasing SFH with random starbursts}
\label{sim:expbur}

In our second set of simulations, we use a prescription similar to
\citet{Chen2009} to create a set of more realistic spectra on which to
test our methods.  The added SFH complexities over the simulations in \S
\ref{sim:exp} are:
\begin{enumerate}
\item The time $t_{form}$ when a galaxy begins star formation is
  distributed uniformly between 0 and 5.7 Gyr after the Big Bang,
  where the Universe is assumed to be 13.7 Gyr old.
\item We again have SFR $\propto \exp(\gamma t)$, but now allow
  $\gamma$ to vary between galaxies.  For each galaxy we draw $\gamma$
  from a uniform distribution between 0.7 and 1 $\textrm{Gyr}^{-1}$.
\item We allow for starbursts with equal probability at all times.
  The probability a starburst begins at time $t$ is normalized so that
  the probability of no starbursts in the life of the galaxy is 33\%.
  The length of each burst is distributed uniformly between 0.03 and
  0.3 Gyr and the fraction of total stellar mass formed in the burst
  in the past 0.5 Gyr is distributed logarithmically between 0 and 0.1.
  The SFR of each starburst is constant throughout the length of the burst.
\end{enumerate}
As above, each galaxy spectrum is generated as a mixture of SSPs of up
to 196 time bins, with a randomly drawn metallicity in each bin.
Allowing for random starbursts over the exponential SFH increases the
importance of younger populations in the simulated galaxies.  We test
our methods on 500 simulated spectra, 250 with S/N=10 and 250 with
S/N=20 at $\lambda_0$.
Across each spectrum, SFH, $\log\langle Z\rangle_L, A_V, \sigma_*$ and
the random noise all vary.  We fit bases of size 45 and 150.

\subsubsection{Results}

\begin{figure}
  \centering
  \epsfig{figure=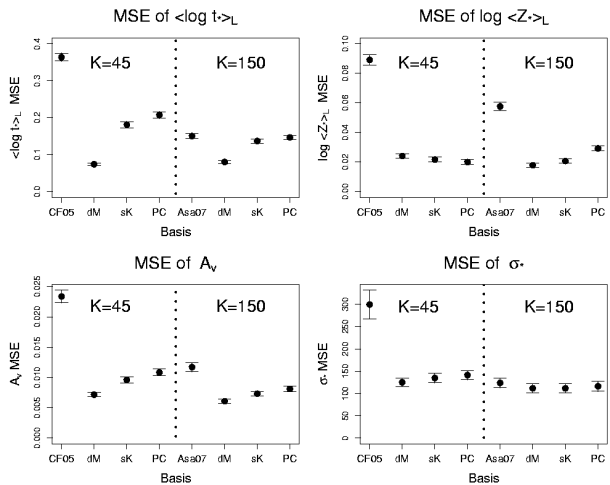,width=80mm}
  \caption{\emph{MSE of four parameter estimates for
   simulated spectra with
      exponentially decreasing SFH and random starbursts.}
  Plotted for each basis is the MSE plus standard error across 500
  simulations.  The $K$-means bases outperform bases from the
  literature for both choices of $K$.  Diffusion $K$-means tends to
  find the best parameter estimates, showing no significant
  change in accuracy between $K$=45 and 150.}
  \label{wild:mse}
\end{figure}

\begin{figure}
  \centering
  \epsfig{figure=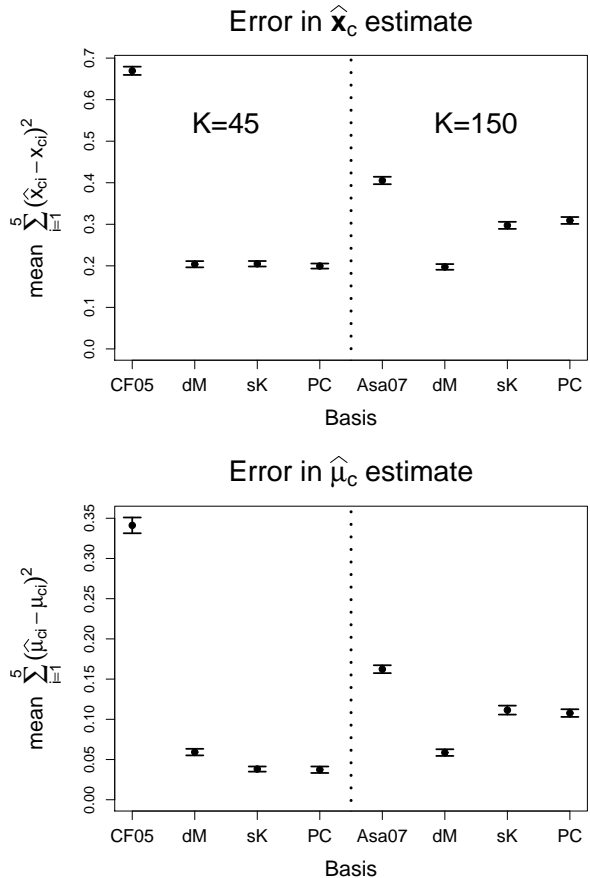,width=80mm}
  \caption{\emph{Error in SFH estimation for simulated spectra with
       exponentially decreasing SFH and random starbursts.}
    Average discrepancies of the estimated condensed light fraction
    ($\mathbf{\widehat{x}}_c$) and condensed mass fraction
    ($\mathbf{\widehat{\mu}}_c$) vectors to the truth, averaged across
    500 simulations.
    Diffusion $K$-means generally produces more accurate SFH estimates than the
    other bases considered.  All $K$-means bases find much more
    accurate estimates than the literature bases.}
  \label{wild:sfh}
\end{figure}

For this set of simulations, we only look at results for the diffusion
$K$-means $\epsilon$ that achieves the 
lowest $\chi^2$ fit error for each $K$; the optimal values are $\epsilon$=250
for the $K=45$ basis and $\epsilon$=500 for the $K=150$ basis.  As
before, results are not very sensitive to the specific choice of
$\epsilon$.  Note that the
$\epsilon$=250 basis of size 45 also achieved good fits to the simpler
simulations in \S \ref{sim:exp}.
In Fig. \ref{wild:mse} we plot the MSE of the physical parameter estimates for
each of the 4 bases for $K$=45 and 150.  The diffusion
$K$-means bases perform significantly better than all other bases for mean
age and $A_V$ estimation and achieve similar results as the other
$K$-means bases for $\log \langle Z_* \rangle_L$ and $\sigma_*$
estimation.  The $K$-means bases heavily outperform the hand-chosen
bases (CF05 \& Asa07) for each parameter.  The same trends are seen
for error in estimation of time-binned star formation histories
(Fig. \ref{wild:sfh}).  The error estimates for all bases, split
into S/N=10 and S/N=20 galaxy spectra, are in Table \ref{wild:tab}.

\setcounter{table}{0}
\begin{table*}
 \begin{minipage}{215mm}
 \caption{Summary of parameter error estimates for simulated spectra with exponentially decreasing SFH and random starbursts.}
 \label{wild:tab}
 \begin{tabular}{ccllllll|llllll}
   \hline
   &     & \multicolumn{6}{c}{S/N=10} &
   \multicolumn{6}{c}{S/N=20}\\
   \hline
   $K$ & Basis & $\langle \log t_* \rangle_L$ & $\log \langle Z_*
   \rangle_L$ & $A_V$ & $\sigma_*$ & $\mathbf{x}_c$ & $\mathbf{\mu}_c$
   & $\langle \log t_* \rangle_L$ & $\log \langle Z_*
   \rangle_L$ & $A_V$ & $\sigma_*$ & $\mathbf{x}_c$ & $\mathbf{\mu}_c$\\
   \hline
&CF05& 0.368& 0.089& 0.025& 369.0& 0.664& 0.349& 0.360& 0.089& 0.022& 229.6& 0.675& 0.333\\
45&dM& {\bf 0.074} &0.024 &{\bf 0.007} &{\bf 156.4} &{\bf 0.195} &0.059 &{\bf 0.074} &0.025 &{\bf 0.007} & {\bf 93.3} &0.213 &0.059\\
&sK &0.194 &0.021 &0.010 &171.7 &0.219 &{\bf 0.046} &0.169 &0.022 &0.009 & 97.3 &0.191 &0.030\\
&PC& 0.222 &{\bf 0.020} &0.012 &174.6 &0.212 &{\bf 0.046} &0.193 &{\bf 0.020} &0.010 &107.6 &{\bf 0.187} &{\bf 0.029}\\
\hline
&Asa07 &0.167 &0.059 &0.013 &169.6 &0.421 &0.175 &0.134 &0.056 &0.010 & 77.4 &0.390 &0.150\\
150&dM &{\bf 0.093} &{\bf 0.018} &{\bf 0.007} &{\bf 153.1} &{\bf 0.213} &{\bf 0.067} &{\bf 0.068} &{\bf 0.018} &{\bf 0.005} & 70.1 &{\bf 0.182} &{\bf 0.050}\\
&sK& 0.150 &0.021 &0.008 &153.4 &0.303 &0.118 &0.124 &0.020 &0.007 & {\bf 69.5} &0.292 &0.105\\
&PC &0.163 &0.029 &0.009 &160.9 &0.317 &0.117 &0.130 &0.029 &0.007 & 71.1 &0.301 &0.099\\
\hline
 \end{tabular}
 \end{minipage}
\end{table*}

An interesting trend appears in Figs. \ref{wild:mse}-\ref{wild:sfh}.
For the hand-chosen bases, we
see considerable reduction
in the estimation error of every parameter in going from the $K$=45 basis of CF05
to the $K$=150 basis of Asa07.  However, in the $K$-means bases, and
particularly for the diffusion $K$-means basis, there is no
substantial lowering of the MSE of any parameter in increasing from $K$=45
to $K$=150.  This tells us that even with as few as 45 prototypes,
diffusion $K$-means provides us a basis that can successfully fit
these particular simulations.  Moving to $K$=150 prototypes allows us
no significant gains in parameter estimation while costing us $\sim$10
times the computational time.  Note, however, that although a basis of
size 45 may be sufficient for this set of simulations, a set of real galaxy
spectra may be more complicated and require a larger basis of prototypes to
achieve good fits.

\begin{figure}
  \centering
  \epsfig{figure=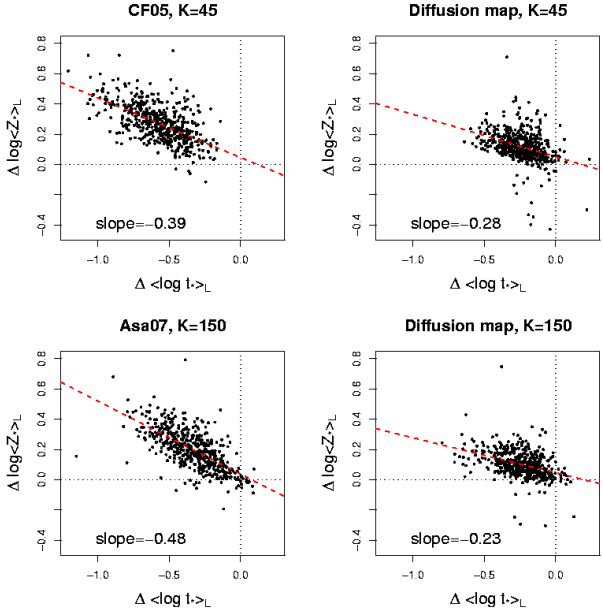,width=80mm}
  \caption{\emph{Age-metallicity degeneracy in parameter estimation for simulated spectra with
       exponentially decreasing SFH and random starbursts.}  Diffusion
   $K$-means bases show significantly less degeneracy in parameter estimates than the
   literature bases, shown by a smaller slope in the linear trend.
   Diffusion $K$-means bases also produce less biased estimates of
   both parameters.}
  \label{wild:ageZ}
\end{figure}

A common difficulty in population synthesis studies is the so-called
age-metallicity degeneracy, where young, metal-rich SSPs are confused
with old, metal-poor SSPs.   Using a regular grid of SSP prototypes in
population synthesis can exacerbate this problem because several
SSPs with nearly identical spectra (but different ages and metallicities) will likely 
be included, inducing degeneracy in the spectral model (\ref{cfmodel}), 
and causing the estimated SFH parameters for each galaxy to be highly 
variable.  A diffusion
$K$-means basis, on the other hand, picks SSP prototypes whose spectra
more evenly span the range of spectral characteristics, minimizing the
inclusion of very similar spectra and thus decreasing the degeneracy
effects in the minimization of (\ref{chisq}).
In Fig. \ref{wild:ageZ} we see that the age-Z degeneracy
is much smaller in diffusion $K$-means bases, quantified by the slope
of the linear trend in $\Delta \log\langle Z_* \rangle_L$ versus $\Delta
\langle \log t_* \rangle_L$.  Some degeneracy in age-metallicity still
exists for the diffusion $K$-means bases, but the magnitude of the
slope of the trend is half.

One may note that both $\log\langle Z_*
\rangle_L$ and $\langle \log t_* \rangle_L$ are significantly biased
for all 4 bases in Fig. \ref{wild:ageZ}.
This probably occurs due to the simulation prescription, which
includes mostly old SSPs from a large, fine grid of age and Z.  One
may be inclined to use these estimated biases to correct the estimates
of real galaxy spectra.  However, the amount of bias is dependent on
the particular prescription of SFH in the simulations.  Since there is no
general model that describes galaxy SFH for a wide range of galaxies,
there is no straightforward way to estimate the bias, if any, that
will occur.

\section{Application to SDSS spectra}
\label{sdss}

In this section, we apply the fitting techniques to a sample of SDSS spectra to
compare the results using different bases. We compare the physical
parameters estimated from each of three different bases: Asa07,
diffusion map with $K$=45, and diffusion map with $K$=150.  We also
compare parameter estimates to those estimated with the MPA/JHU data
base.  Our goal in this section is to study how the choice of basis
affects the final parameter estimates for a large set of real galaxy
spectra.

\subsection{Data preparation}

Our data sample consists of spectra 
from ten arbitrarily chosen spectroscopic plates of SDSS DR6
(0266$-$0274 inclusive, and 0286; \citealt{Adelman2008}) that are
classified as galaxy spectra.  The goal is to test our methods on a
sample that is large enough that we include galaxies with a broad
range of star formation and metallicity histories but not so large
that analysis is computationally cumbersome. In a future work, we will
present results for the more than 700k galaxy spectra in SDSS DR7.

We remove spectra from this sample by applying three cuts.  The first,
motivated by aperture considerations, is to analyse only spectra
whose SDSS redshift estimates $\geq$ 0.05.  
Second, we determine the proportion of the 3850 wavelength bins that
are flagged
as bad.  If this proportion exceeds 10\%, we remove the spectrum from the
sample; if not, we retain the spectrum for further analysis.  
Third, we select galaxy spectra with a SDSS redshift confidence level
$\geq$ 0.35.  These cuts leave us with a sample of 3046 galaxy spectra.

We bring all spectra to rest frame using the SDSS redshift estimates.
In the fitting routines, we exclude any spectral bins with SDSS flag
$\geq$ 2.  This eliminates any sky residuals, bad pixels, and artefacts
from the data reduction.  We additionally
mask the pixels in the regions within and around emission lines as in
\citet{CF2005}.  
Results of the STARLIGHT spectral fits give a mean
$\chi^2/N_{\lambda}$ of 1.03 for each of the three bases used, close to a
value of 1 that would be expected if wavelength bin fluxes were
independently distributed Gaussian random variables.  Note that
this value varies considerably from the mean value of 0.78 found in
the analysis of 50362 SDSS spectra by \citet{CF2005}.  This difference
can be attributed to our careful handling of spectral bin error estimates
in our rebinning of the spectra to a dispersion of 1 measurement per \AA.

\subsection{Parameter estimates}

\begin{figure}
  \centering
  \epsfig{figure=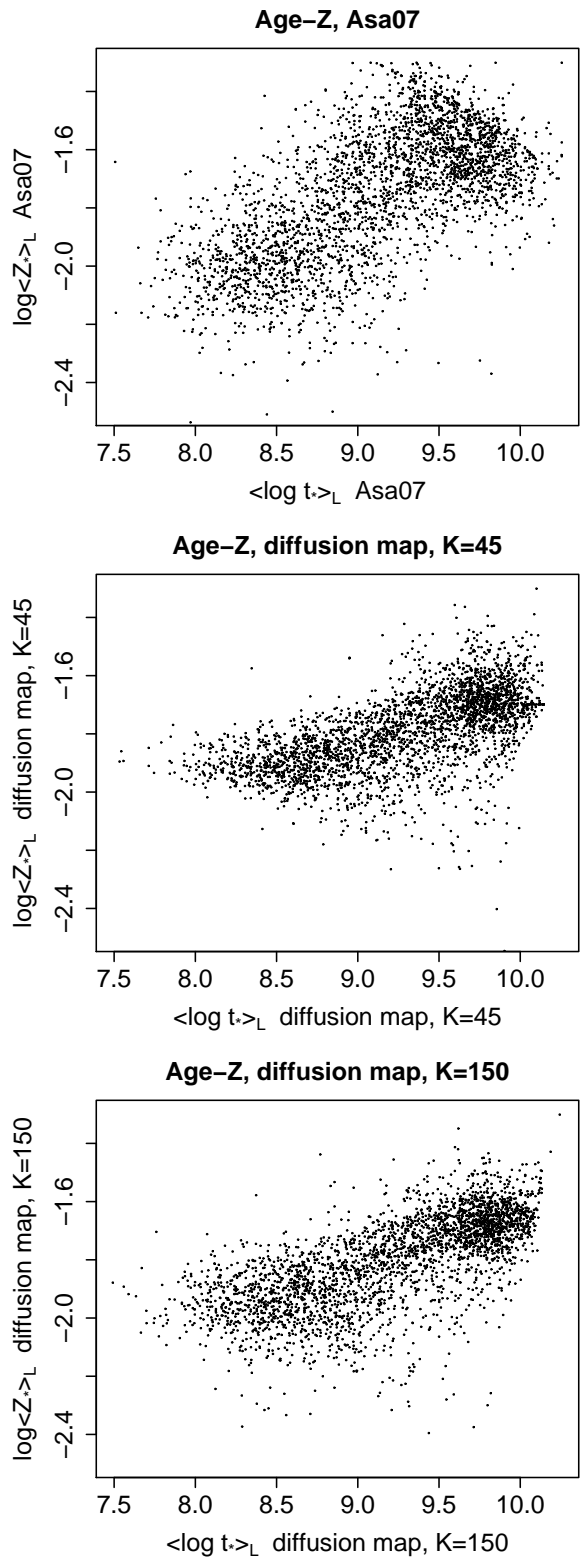,width=80mm}
  \caption{\emph{Estimated ages and metallicities for 3046 SDSS
      galaxies.}  Diffusion $K$-means estimates tend to lie tightly
    around a linear trend in this plane while Asa07 estimates show
    large dispersion in the perpendicular direction, which corresponds
  to the direction of age-Z degeneracy in galaxy population synthesis studies.}
  \label{sdss:tZ}
\end{figure}

The STARLIGHT fitting routine is used to fit all 3046 SDSS spectra
with each of the three bases.  Each basis yields
parameter estimates that are substantially different.  In Fig.
\ref{sdss:tZ} we plot each galaxy's $\log \langle Z_* \rangle_L$ versus $\langle
\log t_* \rangle_L$ estimate for each basis.  There are several
notable differences in the $\log \langle Z_* \rangle_L-\langle
\log t_* \rangle_L$ plane for each basis.  First, both diffusion map
bases produce estimates that are tightly centered around an increasing
linear trend while the Asa07 estimates are diffusely spread around
such a trend.  Moreover, the direction of discrepancy in the Asa07
estimates from the linear trend corresponds exactly with the direction
of degeneracy between old, metal-poor and young, metal-rich galaxies.
This suggests that the observed variability along this direction is
not due to the physics of these galaxies, but rather is caused by
confusion stemming from the choice of basis.

Secondly, the slopes of the
estimated trends produced by the diffusion map bases (0.146 for $K$=45
and 0.180 for $K$=150) are considerably
smaller than the slope of the trend in the Asa07 estimates (0.250).
These estimates have implications for the physical models that govern
galaxy evolution.  Finally, the $K$=45 diffusion map basis estimates
no young, metal-poor galaxies, showing a sharp decline from
densely-populated regions of the parameter space to vacant regions.
This suggests that this small number of prototypes is not sufficient to
cover the parameter space; particularly, young, metal-poor SSPs have
been neglected in the basis.

\begin{figure}
  \centering
  \epsfig{figure=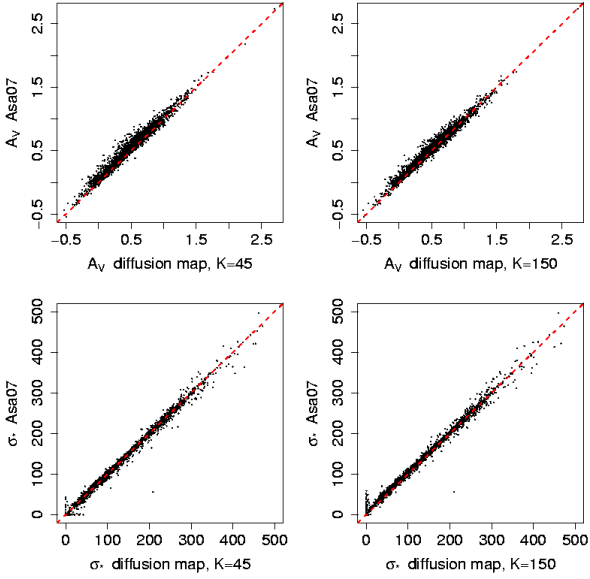,width=80mm}
  \caption{\emph{Estimated $A_V$ and $\sigma_*$ parameters for 3046
      SDSS galaxies.}  The Asa07 $A_V$ estimates tend to be larger
    than the corresponding diffusion map estimates.  The
    $\sigma_*$ estimates agree for all 3 bases.}
  \label{sdss:kin}
\end{figure}

Next we compare kinematic parameter estimation.  Fig. \ref{sdss:kin}
shows that the Asa07 basis tends to consistently estimate slightly
larger values of $A_V$ than either diffusion map basis.  On average,
Asa07 estimates of $A_V$ are 0.045 larger than diffusion map $K$=45
estimates and 0.024 larger than diffusion map $K$=150 estimates.
Estimates of $\sigma_*$ are consistent between all three bases.

\begin{figure}
  \centering
  \epsfig{figure=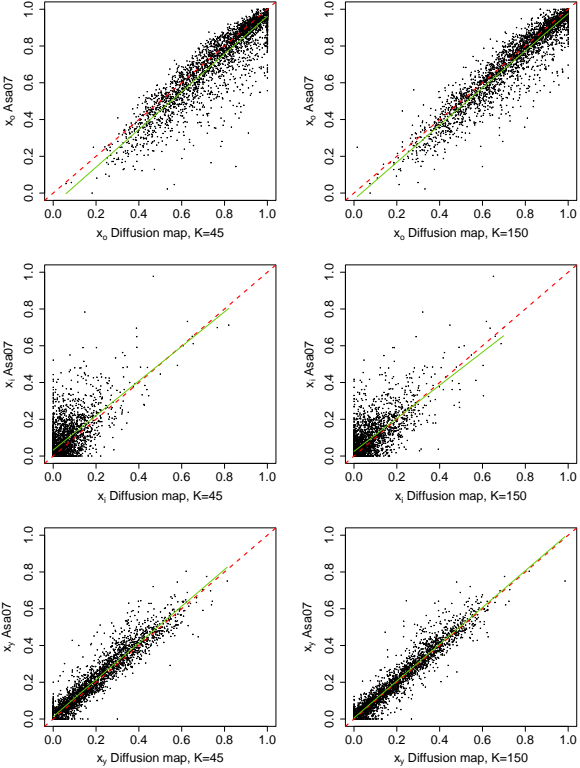,width=80mm}
  \caption{\emph{Age-binned population vector for 3046 SDSS galaxies.}  Plotted are the
    Asa07 estimates of the 3-dimensional age-binned population vector
    versus those of two diffusion $K$-means bases.  Diffusion map
    tends to find slightly smaller contributions of the oldest populations.}
  \label{sdss:sfh}
\end{figure}

Finally, we compare age-binned SFH estimates found by each basis.
Like \citet{CF2005}, we condense the SFH to three parameters: old,
$x_{\textrm o}$ ($t_j>10^9$ yr), intermediate, $x_{\textrm i}$ ($10^8\le t_j \le 10^9$
yr), and young, $x_{\textrm y}$ ($t_j<10^8$ yr).  Plots of diffusion map
estimates of each component versus Asa07 estimates are in Fig.
\ref{sdss:sfh}.  The estimated SFHs show a high degree of
correspondence between the bases.  On average, each diffusion map basis
obtains slightly higher estimates of $x_{\textrm o}$ and slightly lower
estimates of both $x_{\textrm i}$ and $x_{\textrm y}$.

\subsection{Comparisons to the MPA/JHU data base}

We compare the SDSS parameters estimated with the
three different choices of STARLIGHT basis with parameters
obtained using measured emission line and continua values from the
MPA/JHU data base.  The MPA/JHU collaboration has made their SDSS
spectral measurements publicly available on the MPA SDSS
website\footnote{{\tt http://www.mpa-garching.mpg.de/SDSS/}}
(\citealt{K03,Bri2004,Tre2004}).  Although we do not expect any of
the basis choices to produce estimates that perfectly correspond to
the independent MPA/JHU estimates, we analyse the relationships
between the estimates as a further test of the accuracy of each basis.

\subsubsection{$A_V$ and $A_V^{\rm neb}$}

We use the Balmer emission-line ratio $H\alpha / H\beta$ to estimate
the nebular extinction, $A_V^{\rm neb}$ , of each galaxy.  Following
\citet{Ost1989}, we
assume case B recombination, a density of 100 cm$^{-3}$ and a
temperature of 10$^4$ K to compute $A_V^{\rm neb}$ according to
(\ref{Aneb1})-(\ref{Aneb2}):
\begin{eqnarray}
  \label{Aneb1}
  A_V^{\rm neb} &=& c \times 3.1/1.47\\
  \label{Aneb2}
  c &=& -2.70 (2.86 - \log(H\alpha/H\beta)_{\rm obs})
\end{eqnarray}
where we have used the same constants used in \citet{Chen2009}.

\begin{figure}
  \centering
  \epsfig{figure=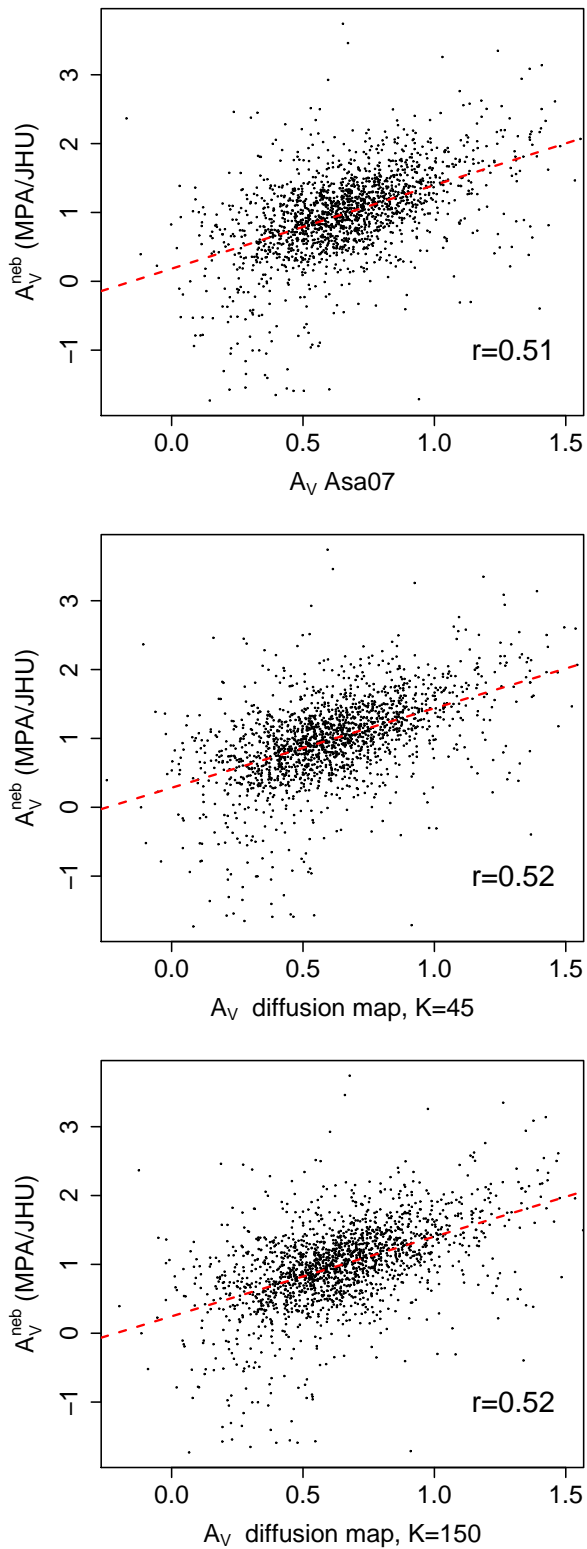,width=80mm}
  \caption{\emph{MPA/JHU versus population synthesis reddening
      estimates for SDSS galaxies.}  All three bases considered find
    estimates that are generally consistent with the independent
    MPA/JHU emission line estimates.}
  \label{jhu:A}
\end{figure}

Fig. \ref{jhu:A} shows plots of the values of $A_V^{\rm neb}$
computed from the MPA/JHU data base versus the values of the stellar
extinction, $A_V$, computed by each of the
three different bases in the STARLIGHT code.  We have only plotted the
1755 SDSS galaxies in
our sample whose $H\alpha$ and $H\beta$ emission line measurements
both have signal-to-noise ratio $\ge 3$.  We find that $A_V^{\rm neb}$ and
$A_V$ are loosely linearly related for each basis.  The estimated
Spearman correlation is 0.51 for the Asa07 basis and 0.52 for each
diffusion map basis.  The slope of the linear trend is slightly higher
for the Asa07 basis (1.21) than for the diffusion map bases (1.16 for
each basis).  All three bases obtain stellar extinction estimates that are
consistent with the nebular extinction values $A_V^{\rm neb}$, and
there are no significant differences between the strength of the
relationships between the estimates of each basis and the nebular
extinction estimates.

\subsubsection{$\langle \log t_* \rangle_L$ and $D_n(4000)$ \&
  $H\delta_A$}

As in \citet{Chen2009}, we compare our spectral synthesis estimates to
two age indicators from the
MPA/JHU database: $D_n(4000)$ (\citealt{Bal1999}) and $H\delta_A$
(\citealt{Wor1997}).  The value $D_n(4000)$ describes the break in the
4000 \AA~ region of the galaxy spectrum; higher values generally
correspond to increasing ages of the galaxy's stellar subpopulations.
Larger strength of the $H\delta$ absorption line, $H\delta_A$, indicates
higher star formation rates in the past 0.1-1 Gyr of a galaxy's
lifetime.  Hence, we
expect our estimates of $\langle \log t_* \rangle_L$ to correlate
positively with $D_n(4000)$ and negatively with $H\delta_A$.

\begin{figure}
  \centering
  \epsfig{figure=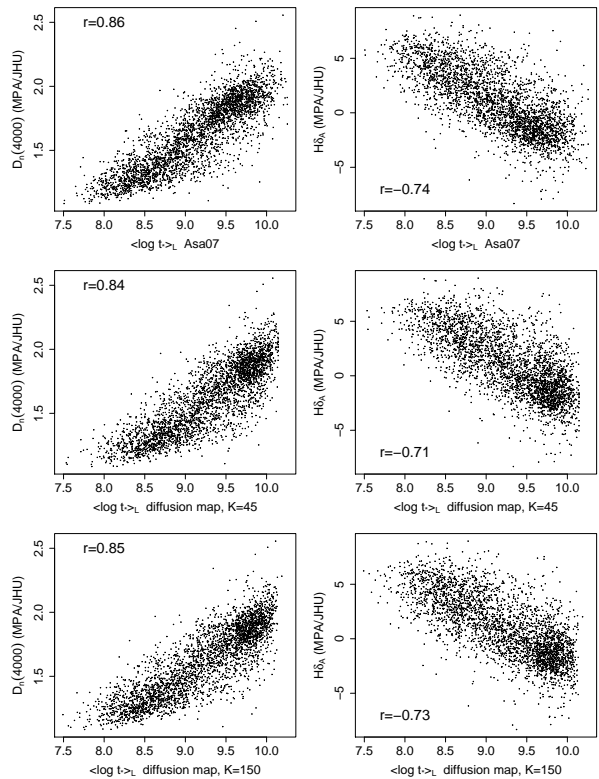,width=80mm}
  \caption{\emph{MPA/JHU versus population synthesis age indicators
      for SDSS galaxies.}  All three bases considered retrieve
    estimates that are consistent with two different MPA/JHU
    indicators of age of stellar population.}
  \label{jhu:t}
\end{figure}

In Fig. \ref{jhu:t}, we see that the $\langle \log t_* \rangle_L$
estimates show the expected relationships with both $D_n(4000)$ and
$H\delta_A$ for each of the three bases.  Correlations are strongly
positive for the $D_n(4000)-\langle \log t_* \rangle_L$ trend and
strongly negative for the $H\delta_A-\langle \log t_* \rangle_L$
trend. We have plotted the 3031 galaxies in our sample that have 
reliable $D_n(4000)$ and $H\delta_A$ measurements in the MPA/JHU catalogue. There
is no significant
difference in each relation between the bases.

\subsubsection{$\langle Z_* \rangle_M$ and $Z_{\rm neb}^{\rm o3n2}$}

Lastly, we compare our STARLIGHT estimates of $\langle Z_* \rangle_M$
with a measure of nebular metallicity based on the ratio of emission lines
[OIII]$\lambda 5007$ and [NII]$\lambda 6583$, provided by MPA/JHU. 
Particularly, the quantity $Z_{\rm neb}$ measures the oxygen abundance
of a galaxy, and is defined as
\begin{equation}
  \label{eq:Zneb}
  Z_{\rm neb} = -3.45-.25\times {\rm O}_3{\rm N}_2
\end{equation}
where O$_3$N$_2 =
\log([{\rm OIII}]\lambda 5007 / [{\rm NII}]\lambda 6583)$ (\citealt{Pet2004}).
 All
emission line fluxes are first corrected for extinction using the $A_V$
estimates found in STARLIGHT fits.  We would expect $\langle Z_*
\rangle_M$ to be positively related to 
$Z_{\rm neb}$, as the former measures stellar metallicity and the
latter measures nebular oxygen abundance.

\begin{figure}
  \centering
  \epsfig{figure=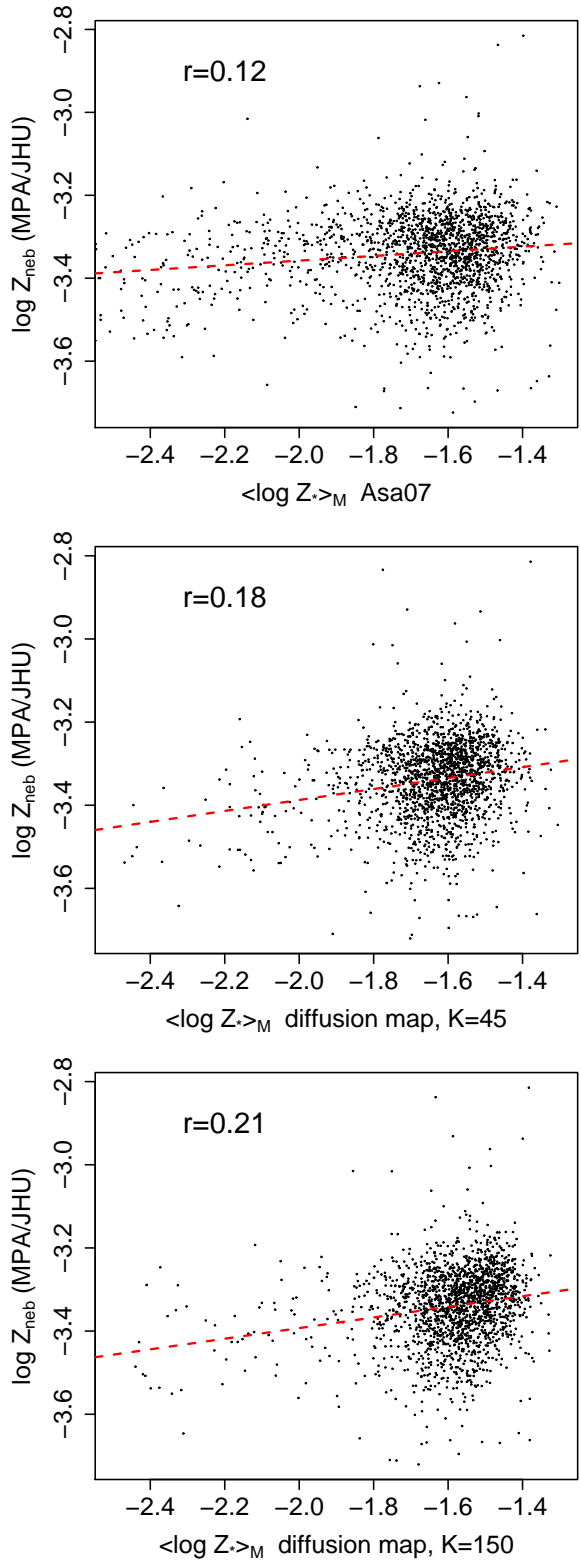,width=80mm}
  \caption{\emph{MPA/JHU versus population synthesis metallicity
      estimates for SDSS galaxies.}  MPA/JHU estimates of nebular
    oxygen abundance correlate more strongly with diffusion $K$-means
    estimates of stellar metallicity than the Asa07 estimates.}
  \label{jhu:Z}
\end{figure}

In Fig. \ref{jhu:Z} we see that all three bases find $\langle
Z_* \rangle_M$ estimates that correlate positively with $Z_{\rm neb}$,
and that the $Z$=150 diffusion map basis estimates correlate more
strongly with $Z_{\rm neb}$, with Spearman correlation coefficient of
0.21, compared to the Asa07 basis estimates, which have a correlation of 0.12
with the $Z_{\rm neb}$ estimates.  Furthermore, the best-fit linear
trend of the $Z_{\rm neb}$ versus diffusion map $K$=150 estimates has
a significantly higher slope, 0.128 than the linear trend of $Z_{\rm
  neb}$ vs. Asa07 basis, 0.056.  This result suggests that the
diffusion map basis obtains more reliable metallicity estimates than
the Asa07 basis.

\section{Discussion}

We have presented a method to choose an appropriate basis of SSP
prototypes to be used in population synthesis models of observed
galaxy spectra.  Particularly, we have used the STARLIGHT fitting
routines of \citet{CF2005} to study the ability of both
literature and diffusion $K$-means bases to model both simulated
and SDSS galaxy spectra.  On simulated galaxies, our
diffusion $K$-means basis outperforms hand-chosen literature
bases as well as bases constructed from standard $K$-means
or PC $K$-means.  For real SDSS galaxy spectra, we
show that the diffusion $K$-means basis finds parameter estimates that
differ substantially from the estimates retrieved by the basis of
\citet{Asa2007}.  The diffusion $K$-means estimates seem to alleviate the
age-metallicity degeneracy that has been detrimental to previous
population synthesis studies of large data bases of galaxies.

At this point, more discussion of the age-metallicity degeneracy is
warranted.  Degeneracy arises because several SSPs
have very similar spectral characteristics, but different ages and metallicities.
When degenerate SSPs are included, estimates can be highly
variable. This is shown in the first panel in Fig. \ref{sdss:tZ} and in the
STARLIGHT code itself, which in the presence of degeneracy tends to 
converge to different solutions due to the random walk taken through the parameter
 space by the Metropolis algorithm.  The diffusion $K$-means algorithm,
on the other hand, includes only the mean spectrum and mean parameters
of each set of degenerate spectra.  This choice minimizes the statistical
risk of the final solution by decreasing the variance and bias of the estimates
(our measure of statistical risk is the MSE,
which is the sum of the variance and squared bias of an estimate), as seen
in simulations (Figs. 5-9).  Our SFH estimates for SDSS spectra are significantly less
variable in the direction of the age-Z degeneracy, as seen in Fig. \ref{sdss:tZ},
while the relative amount of bias is impossible to quantify without knowledge
of the true SFH of each galaxy.

 In the face of degeneracies and with no other outside information available,
our proposed method (of picking the mean of several degenerate solutions)
is the best one can do in terms of minimizing the statistical risk of the estimates.
However, with more information about what features in the SSP spectra
are most strongly related to their ages and metallicities, we could potentially improve our
estimates by incorporating such information
in the construction of the similarity measure
$s(\mathbf{b}_i,\mathbf{b}_j)$ in (3).  The weight matrix produced by an expert-produced
similarity measure
may produce a more complete coverage of the parameter space
in our SSP prototypes and help to remove degeneracies in the parameter estimates.
Further study is required in this direction.

The ultimate goal of this project is to be able to quickly and accurately estimate
physical property and SFH parameters for a large data base of
galaxies.  As an extension of this work, we plan to study how to
quickly and effectively extend estimates found for a small set of galaxies using the
computationally intensive STARLIGHT routine to a large database of
galaxies.  To achieve this goal, we must exploit the
low-dimensional structure of these data.  This
approach was taken by \citet{Rich2009} in galaxy redshift estimation.
These methods must be further developed and refined to reliably
estimate SFH parameters.

The methods presented in this paper are applicable in a wide range of
problems in astrophysical data analysis.  Diffusion map is a powerful
tool to uncover simple structure in complicated, high-dimensional
data, whether they be spectra, photometric colours, two-dimensional
images, data cubes, etc.  Specifically, diffusion map is both more
flexible and more widely applicable than PCA, which relies on the
linear correlation structure of the data and assumes that the data
reside on a low-dimensional, linear manifold.  The diffusion $K$-means
method can
be used in a variety of problems often encountered in astrophysics
where a large set of observed or model data are described with a few
prototype examples.

Future work will attempt to more concretely define criteria
for the performance of algorithms that define $K$
prototypes from $N$ data points and to
theoretically justify the use of diffusion $K$-means.
We plan to explore more deeply the relationship
between the $K$-means methods introduced here and to compare the
methods introduced in this paper to other methods of basis
selection, such as Basis Pursuit and Matching Pursuit.
We also will investigate how meaningful standard errors
can be attached to our parameter estimates using any choice of basis.

\section*{Acknowledgments}

We thank the reviewer for their helpful comments.

This work was supported by NSF grants CCF-0625879 and DMS-0707059 and ONR grant N00014-08-1-0673.

Funding for the SDSS and SDSS-II has been provided by the Alfred P. Sloan Foundation, the Participating Institutions, the National Science Foundation, the U.S. Department of Energy, the National Aeronautics and Space Administration, the Japanese Monbukagakusho, the Max Planck Society, and the Higher Education Funding Council for England. The SDSS Web site is http://www.sdss.org/. The SDSS is managed by the Astrophysical Research Consortium for the Participating Institutions. The Participating Institutions are the American Museum of Natural History, Astrophysical Institute Potsdam, University of Basel, Cambridge University, Case Western Reserve University, University of Chicago, Drexel University, Fermilab, the Institute for Advanced Study, the Japan Participation Group, Johns Hopkins University, the Joint Institute for Nuclear Astrophysics, the Kavli Institute for Particle Astrophysics and Cosmology, the Korean Scientist Group, the Chinese Academy of Sciences (LAMOST), Los Alamos National Laboratory, the Max-Planck-Institute for Astronomy (MPIA), the Max-Planck-Institute for Astrophysics (MPA), New Mexico State University, Ohio State University, University of Pittsburgh, University of Portsmouth, Princeton University, the United States Naval Observatory, and the University of Washington.

The STARLIGHT project is supported by the Brazilian agencies CNPq, CAPES and
FAPESP and by the France-Brazil CAPES/Cofecub program.

\label{lastpage}


\begin{thebibliography}{}
\bibitem[Adelman-McCarthy et al.(2008)]{Adelman2008} Adelman-McCarthy, J.~K., et al.~2008, ApJS, 175, 297
\bibitem[Alongi et al.(1993)]{Alo1993} Alongi, M., Bertelli, G.,
  Bressan, A., Chiosi, C., Fagotto, F., Greggio, L., and Nasi,
  E., ~1993, A\&AS, 97, 851
\bibitem[Asari et al.(2007)]{Asa2007} {Asari}, N.~V., {Cid Fernandes}, R., {Stasi{\'n}ska}, G.,	{Torres-Papaqui}, J.~P., {Mateus}, A., {Sodr{\'e}}, L.,	{Schoenell}, W., and {Gomes}, J.~M., ~2007, MNRAS, 381, 263
\bibitem[Balogh et al.(1999)]{Bal1999} Balogh, M.~L., Morris, S.~L.,
  Yee, H.~K.~C., Carlberg, R.~G. and Ellingson, E., ~1999, ApJ, 527, 54
\bibitem[Bica(1988)]{Bica1988} Bica, E., ~1988, A\&A, 195, 76
\bibitem[Bressan et al.(1993)]{Bre1993} Bressan, A., Fagotto, F.,
  Bertelli, G., Chiosi, C., ~1993, A\&AS, 100, 647
\bibitem[Brinchmann et al.(2004)]{Bri2004} {Brinchmann}, J.,
  {Charlot}, S., {White}, S.~D.~M., {Tremonti}, C., {Kauffmann}, G., {Heckman}, T., and {Brinkmann}, J., ~2004, MNRAS, 351, 1151
\bibitem[Bruzual \& Charlot(2003)]{BC03} Bruzual, G., and Charlot,
  S., ~2003, MNRAS, 344, 1000
\bibitem[Cardelli, Clayton \& Mathis(1989)]{CCM1989} Cardelli, J.~A., Clayton,
  G.~C., and Mathis, J.~S., ~1989, ApJ, 345, 245
\bibitem[Chabrier(2003)]{Cha2003} Chabrier, G., ~2003, PASP, 115, 763
\bibitem[Chen et al.(2009)]{Chen2009} Chen, X.~Y., Liang, Y.~C.,
  Hammer, F., Zhao, Y.~H. and Zhong, G.~H., ~2009, A\&A, 495, 457
\bibitem[Cid Fernandes et al.(2001)]{CF2001} Cid Fernandes, R., Sodre,
  L., Schmitt, H.~R., Leao, J.~R.~S., ~2001, MNRAS, 325, 60
\bibitem[Cid Fernandes et al.(2004)]{CF2004} Cid Fernandes, R., Gu,
  Q., Melnick, J., Terlevich, E., Terlevich, R., Kunth, D., Rodrigues
  Lacerda, R., and Joguet, B., ~2004, MNRAS, 355, 273
\bibitem[Cid Fernandes et al.(2005)]{CF2005} {Cid Fernandes}, R.,
  {Mateus}, A., {Sodr{\'e}}, L., {Stasi{\'n}ska}, G., and {Gomes}, J.~M., ~2005, MNRAS, 358, 363
\bibitem[Coifman \& Lafon(2006)]{Coif2006} Coifman, R.~R., \& Lafon,
  S., ~2006, Appl. Comput. Harmon. Anal., 21, 5
\bibitem[Fagotto et al.(1994a)]{Fag1994a} Fagotto, F., Bressan, A.,
  Bertelli, G., Chiosi, C., 1994a, A\&AS, 104, 365
\bibitem[Fagotto et al.(1994b)]{Fag1994b} Fagotto, F., Bressan, A.,
  Bertelli, G., Chiosi, C., 1994b, A\&AS, 105, 29
\bibitem[Freeman et al.(2009)]{Free2009} Freeman, P.~E., Newman, J.~A.,
  Lee, A.~B., Richards, J.~W., and Schafer, C.~M., in preparation.
\bibitem[Gallazzi et al.(2005)]{Gal2005} Gallazzi, A., Charlot, S.,
  Brinchmann, J., White, S.~D.~M., and Tremonti, C.~A., ~2005, MNRAS, 362, 41
\bibitem[Girardi et al.(1996)]{Gir1996} Girardi, L., Bressan, A.,
  Chiosi, C., Bertelli, G., and Nasi, E., ~1996, A\&AS, 117, 113
\bibitem[Glazebrook et al.(2003)]{Gla2003} Glazebrook, K., et al., ~2003, ApJ, 587, 55
\bibitem[Kauffmann et al.(2003)]{K03} Kauffmann, G. et
  al., ~2003, MNRAS, 341, 33
\bibitem[Lafon \& Lee(2006)]{LafonLee2006} Lafon, S., \& Lee, A., ~2006, IEEE Trans. Pattern Anal. and Mach. Intel., 28, 1393
\bibitem[Le Borgne et al.(2003)]{LeB2003} Le Borgne, J.-F., et al.,
  ~2003, A\&A, 402, 433
\bibitem[Leitherer et al.(1995)]{Leith95} Leitherer, C., Robert, C.,
  Heckman, T.~M., ~1995, ApJS, 99, 173
\bibitem[Mas-Hesse \& Kunth(1991)]{Mas91} Mas-Hesse, J.~M.\& Kunth, D., ~1991, A\&AS, 88, 399
\bibitem[Mathis, Charlot \& Brinshmann(2006)]{Mat2006} Mathis, H.,
  Charlot, S., and Brinshmann, J., ~2006, MNRAS, 365, 385
\bibitem[Moultaka et al.(2004)]{Mou2004} Moultaka, J., Boisson, C.,
  Joly, M., Pelat, D., ~2004, A\&A, 420, 459	
\bibitem[Ocvirk et al.(2006)]{Ocv2006} Ocvirk, P., Pichon, C., Lancon,
  A., and Thiebaut, E., ~2006, MNRAS, 365, 46
\bibitem[Osterbrock(1989)]{Ost1989} Osterbrock, D.~E., ~1989,
  Astrophysics of Gaseous Nebulae and Active Galactic Nuclei (Mill
  Valley: University Science Books)
\bibitem[Panter, Heavens \& Jimenez(2003)]{Pan2003} Panter, B.,
  Heavens, A.~F., and Jimenez, R., ~2003, MNRAS, 343, 1145
\bibitem[Panter et al.(2007)]{Pan2007} Panter, B.,  Jimenez, R.,
  Heavens, A.~F. and Charlot, S., ~2007, MNRAS, 378, 1550
\bibitem[Pelat(1997)]{Pel1977} Pelat, D., ~1997, MNRAS, 284, 365
\bibitem[Pettini \& Pagel(2004)]{Pet2004} Pettini, M. and Pagel,
  B.~E.~J., 2004, MNRAS, 348, L59
\bibitem[Reichardt, Jimenez \& Heavens(2001)]{Rei2001} Reichardt, C.,
  Jimenez, R., and Heavens, A.~F., ~2001, MNRAS, 327, 849
\bibitem[Richards et al.(2009)]{Rich2009} Richards, J.~W., Freeman,
  P.~E., Lee, A.~B., Schafer, C.~M., ~2009, ApJ, 691, 32
\bibitem[Ronen, Arag\'on-Salamanca \& Lahav(1999)]{Ron1999} Ronen, S.,
  Arag\'on-Salamanca, A., and Lahav, O., ~1999, MNRAS, 303, 284
\bibitem[Strauss et al.(2002)]{Str2002} Strauss, M.~A. et al., ~2002, AJ, 124, 1810
\bibitem[Tremonti(2003)]{Tre2003} Tremonti, C., ~2003, PhD
  thesis, Johns Hopkins Univ.
\bibitem[Tremonti et al.(2004)]{Tre2004} Tremonti, C.~A. et al., ~2004, ApJ, 613, 898
\bibitem[Tojeiro et al.(2007)]{toj2007} Tojeiro, R., Heavens, A.~F.,
  Jimenez, R., and Panter, B., ~2007, MNRAS, 381, 1252
\bibitem[Worthey (1994)]{Worth94} Worthey, G., ~1994, ApJS, 94, 687
\bibitem[Worthey \& Ottaviani(1997)]{Wor1997} Worthey, G. and
  Ottaviani, D.~L., ApJS, 111, 377
\bibitem[York et al.(2000)]{York2000} York, D.~G. et al., ~2000, AJ, 120, 1579
\end{thebibliography}
\end{document}